\documentclass[twoside]{article}

\usepackage{PRIMEarxiv}

\usepackage[utf8]{inputenc} 
\usepackage[T1]{fontenc}    
\usepackage{hyperref}       
\usepackage{url}            
\usepackage{booktabs}       
\usepackage{amsfonts}       
\usepackage{nicefrac}       
\usepackage{microtype}      
\usepackage{lipsum}
\usepackage{amsthm}
\usepackage{fancyhdr}       
\usepackage{graphicx}       
\graphicspath{{media/}}     

\usepackage{amsmath,amssymb,amsfonts}
\usepackage{algorithmic}
\usepackage{graphicx}
\usepackage{textcomp}
\usepackage{multirow}
\usepackage{array}
\usepackage{subcaption}
\usepackage{amsmath,nccmath}
\usepackage{amsfonts}
\usepackage{tablefootnote}

\newcommand{\probP}{\text{I\kern-0.15em P}}

\newtheorem{theorem}{Theorem}[section]

\newtheorem{definition}[theorem]{Definition}
\newtheorem{remark}{Remark}

\usepackage{subcaption}
\usepackage{pgfplots}
\pgfplotsset{compat=1.18}
\usepackage{tikz}
\usetikzlibrary{decorations.pathreplacing}
\usetikzlibrary{shapes.geometric, arrows.meta, positioning, calc}

\title{Beyond Distributive Justice: Hermeneutical Fairness in Ad Delivery}

\author{
  Camilla Quaresmini \\
  Politecnico di Milano \\
  Milan, Italy\\
  \texttt{camilla.quaresmini@polimi.it} \\
  \And
  Valentina Breschi \\
  Eindhoven University of Technology \\
  Eindhoven, The Netherlands\\
  \texttt{v.breschi@tue.nl} \\
  \And
  Jessica Leoni \\
  Politecnico di Milano \\
  Milan, Italy\\
  \texttt{jessica.leoni@polimi.it} \\
  \And
  Viola Schiaffonati \\
  Politecnico di Milano \\
  Milan, Italy\\
  \texttt{viola.schiaffonati@polimi.it} \\
  \And
  Mara Tanelli \\
  Politecnico di Milano \\
  Milan, Italy\\
  \texttt{mara.tanelli@polimi.it} \\
  \And
  Giulia De Pasquale \\
  Eindhoven University of Technology \\
  Eindhoven, The Netherlands\\
  \texttt{g.de.pasquale@tue.nl} \\
}

\begin{document}
\maketitle

\begin{abstract}
Fairness in online advertising is often formalized as a distributive justice problem, aiming to ensure that impressions, opportunities, or outcomes are allocated comparably across protected groups. However, even when fairness is ensured according to classical statistical notions, online advertising can still produce systematic harms arising from the ads' content and the way recipients interpret and uptake them. To capture this dimension, we draw on Miranda Fricker's notion of \emph{hermeneutical injustice}. Building on this philosophical account, we model online ad delivery as a mechanism that distributes interpretative resources and can fail in two ways. First, relevant concepts can be withheld through systematic under-exposure, leading to \emph{hermeneutical deprivation}. Second, recipients may experience \emph{hermeneutical distortions} when they are saturated with low uptake or skewed framings. According to these findings, after grounding these mechanisms in exploratory correlational patterns from the \emph{AIDS Advertising Evaluation} surveys (1986-1987), we introduce a group-level hermeneutical fairness constraint and a hermeneutically-aware utility cost in the design of an optimal ad distribution strategy to account for interpretative uptake. We then integrate them into a benchmark, versatile, utility-driven ad allocation framework that already enforces distributive justice, yielding a distributively fair, hermeneutically aware ad allocation framework that prevents the concentration of deprivation and distortion within any protected group. Using controlled simulations, we explore the trade-offs between economic utility, classical distributive fairness constraints (statistical parity and equality of opportunity), and hermeneutical cost and constraints. The results show that a purely utility-based viewpoint drives systematic under-delivery to the disadvantaged group of recipients. Meanwhile, when the hermeneutical stakes of withholding ads are high, distributive constraints can reduce hermeneutical cost at modest utility loss. Conversely, placing a large weight on hermeneutical cost in the absence of distributive constraints can yield ad delivery policies that are highly concentrated on the disadvantaged recipient group. Together, these findings motivate expanding fairness analyses of online advertising beyond classical distributive notions to account for epistemic conditions of interpretation and uptake.
\end{abstract}


\section{Introduction}\label{sec:1}

Algorithmic personalization shapes how people access knowledge and are exposed to online advertising \cite{bozdag2013bias,cotter2020algorithmic,sweeney2013discrimination}. Indeed, search engines, recommender systems, and ad platforms tailor results and sponsored content to inferred preferences \cite{noble2018algorithms,pariser2011filter,kant2020making}, thereby influencing not only what people consume but also what they are able to know \cite{cotter2020algorithmic,harris2022real,shin2025automating}. Since pattern-based personalization is central to modern informational infrastructures, questions about fairness in access to—and the quality of—information in ad delivery are increasingly pressing \cite{deldjoo2022fairness,chien2023fairness,goodman2014informational}.

Empirical work already documents discriminatory patterns across ad domains, including job postings \cite{imana2021auditing,stauffer2018only}, insurance \cite{fabris2021algorithmic}, healthcare \cite{schenker2014ethics}, political campaigns \cite{bar2024systematic}, and societal discourse \cite{ali2021ad}. This evidence has motivated a view on fairness that largely emphasizes distributive concerns, thus prompting its operationalization through quantitative criteria on how benefits, opportunities, or resources (concretely impressions, bids, or outcomes) are allocated across protected groups \cite{baumann2024fairness,datta2018discrimination,watts2021fairness,speicher2018potential,lambrecht2019algorithmic,navon2023bidding,sc_gdp_nl_aas_fd}. While this exposure-focused framing is natural in a setting where delivery is explicitly conditioned on user features, limiting or constraining targeting options does not fully mitigate discriminatory effects \cite{baumann2024fairness,ali2019discrimination,imana2021auditing,sapiezynski2024use}. In particular, \cite{soares2019asymmetric,munoz2024quantifying,phillips2023high,celis2019controlling,bozdag2015breaking,ribeiro2018media,cagle2023using,guo2022lpfs,tang2022together,ross2022echo} show that feature-based targeting and personalization can generate informational asymmetries and contribute to opinion clustering, as users are steered into increasingly segmented information environments. However, even when distributional fairness holds, advertising can still disadvantage groups due to its conceptual content and the interpretative frames it embeds. Indeed, since algorithmic systems mediate the diffusion and validation of knowledge, they shape classificatory boundaries that structure how the world is interpreted \cite{alvarado2023ai,symons2022epistemic,mollema2025taxonomy,hyzen2025epistemic}. This aspect is significant as it proves that currently employed approaches to ensuring fair advertising across domains, from healthcare to employment, remain susceptible to conceptual availability and distortion, which, in turn, can influence how individuals interpret their circumstances and perceived opportunities \cite{gaucher2011evidence,mintzes2002direct,applequist2018updated,ventola2011direct,vakratsas1999advertising}. These considerations emphasize the importance to explicitly account for an \emph{epistemic} perspective - understood broadly as the conditions, limits, and opportunities for knowing - along with distributive accounts in online ads to capture the full social stakes of algorithmic systems, as argued by a growing body of work (see, e.g., \cite{hoffmann2019fairness,selbst2019fairness,shelby2023sociotechnical,wegner2024beyond,grgic2018beyond}).

To theoretically ground our intuition, we draw on Miranda Fricker's framework which distinguishes testimonial from hermeneutical injustice \cite{fricker2007epistemic}.\footnote{This work analyzes how identity power shapes what is intelligible and socially recognized.} These are two forms of epistemic injustice: the former concerns credibility deficits driven by prejudice, while the latter concerns gaps in shared interpretative resources that make experiences difficult to understand In this work we focus on interpretative resources and treat hermeneutical injustice - wrongful gaps in shared interpretative resources that make some experiences unintelligible - as a specific form of the broader category of epistemic injustice. More precisely, we treat epistemic harm as the broader category of wrongs done to subjects in their capacity as knowers. Such harm takes a specifically hermeneutical form when ad delivery shapes access to the interpretative resources needed to render socially significant experiences intelligible. Accordingly, we read targeted advertising not only as a problem of ad allocation, but also as a problem of distributing a specific type of epistemic resources, in particular interpretative resources. Therefore, in this work, we identify two mechanisms through which targeted ads can generate \emph{epistemic harms}: \emph{systematic under-exposure} and \emph{systematic over-exposure}.\footnote{Here, \emph{epistemic harms} are wrongs done to subjects in their capacity \emph{as knowers}.} These can in turn give rise to \emph{hermeneutical harms}: the former to \emph{hermeneutical deprivations}, as relevant concepts are withheld from particular audiences, and the latter to \emph{hermeneutical distortions} through repeated exposure to simplified or skewed framings. We show that these mechanisms are consistent with exploratory patterns found in our analysis of the \emph{AIDS Advertising Evaluation} surveys\footnote{\url{https://discovery.nationalarchives.gov.uk/details/r/C11521962}} (1986-1987). We wish to stress that, although our exploratory empirical illustration relies on a healthcare advertising dataset, the framework we develop is not specific to healthcare. We use the AIDS case as a concrete baseline for motivating and operationalizing a more general account of hermeneutical fairness in ad delivery. This empirical analysis motivates our development of a novel quantitative operationalization of hermeneutical fairness\footnote{We use \emph{justice} in its philosophical sense to denote the relevant normative wrongs in epistemic relations. Our contribution, however, is framed in terms of \emph{fairness} as an evaluative property of the distributive process governing an allocation. On our usage, fairness concerns whether the allocation rule yields just outcomes, where \emph{justice} is assessed at the level of the resulting distributions rather than as a complete account of the underlying wrong.} in terms of costs and constraints that, combined with utility-driven losses and more classical distributive justice constraints (i.e., enforcing statistical parity and equality of opportunity), can be used to design distributionally and hermeneutically fair ad delivery policies. We do so by extending the fair utility-maximizing allocation strategy proposed in \cite{baumann2024fairness} with our proposed hermeneutical cost and constraints, exploring the trade-offs between economic utility, distributive fairness, and hermeneutical fairness via numerical simulations. In the following, we use \emph{hermeneutically-aware} and \emph{hermeneutically fair} to mark two distinct notions. Hermeneutically-aware denotes designs that explicitly model and respond to hermeneutical considerations, whereas hermeneutically fair denotes the stronger condition of satisfying a formal hermeneutical fairness criterion.

The paper is structured as follows. The overall framework of our work is developed in Section~\ref{sec:2}, situating our contribution within the literature on fairness in advertising and motivating why a purely distributive, exposure-based view is incomplete for capturing epistemic harms. Section~\ref{sec:3} presents exploratory empirical evidence consistent with epistemic harms in the  \emph{AIDS Advertising Evaluation} surveys. In Section~\ref{sec:4}, we introduce our formalization of hermeneutical harms and fairness, that we use in Section~\ref{sec:5} to present a utility-maximizing allocation strategy under hermeneutical and classical distributive constraints. Section~\ref{sec:6} presents a set of numerical results aimed at underpinning how group-structured asymmetries interact with standard fairness constraints. Section \ref{sec:discussion} discusses these findings, clarifies the scope of our work and poses awareness on its limitations. Finally, Section \ref{sec:conclusions} concludes with implications for auditing and platform design and highlights future research directions.


\section{From Distributive to Hermeneutical Fairness}\label{sec:2}
Most works on fairness in advertising revolve around distributive concerns. Hence, they tend to primarily focus on how ad delivery, exposure, or outcomes are allocated across protected groups, treating fairness solely as a property of group-level exposure or outcome distributions. This choice is indeed supported by several empirical studies (see, e.g., \cite{datta2018discrimination,sweeney2013discrimination,lambrecht2019algorithmic,ali2019discrimination,imana2024auditing}), which show that targeting and auction dynamics can cause systematic disparities in access to opportunities, including (but not limited to) employment, housing, credit, and education. Correspondingly, audit studies generally compare the delivery of specific ad types across user groups or profiles \cite{imana2021auditing,speicher2018potential,kingsley2020auditing,asplund2020auditing}. Meanwhile, mitigation strategies typically impose group-conditioned constraints on reach/impressions \cite{speicher2018potential,baumann2024fairness,schoeffer2023online,singh2018fairness,timmaraju2023towards,li2025online,molina2023trading}, modify auction or pacing mechanisms \cite{celis2019toward,dalenberg2018preventing,lee2013real,yuan2013real,yuan2025fair}, or target parity in downstream outcomes \cite{watts2021fairness,ali2023problematic,peysakhovich2023implementing}.

However, not all harms in advertising boil down to who sees an ad. Indeed, even with equalized exposure, advertising content can encode normative frames that shape in different ways the identities and options that appear legitimate to different groups of recipients \cite{eisend2010meta,antoniou2020gender,sierra2012ethnic,rossner2023ethnic,davis2018selling,rona2023representation}. For instance, Sony's Dutch billboard for the white PlayStation Portable - which paired the tagline "White is coming'' with racialized imagery - was criticized as racist and withdrawn from the market\footnote{See, e.g., \url{https://www.campaignlive.co.uk/article/sony-withdraws-potentially-racist-dutch-billboards/569738}}. Likewise, the Volkswagen e-Golf advert banned by the UK Advertising Standards Authority\footnote{\url{https://www.theguardian.com/media/2019/aug/14/first-ads-banned-for-contravening-gender-stereotyping-rules}}  relied on gendered role juxtaposition that reinforced stereotypes about who belongs in technical domains.

These cases highlight that a core harm in advertising lies in the creative's normative framing rather than in unequal delivery, and that exposure parity can coexist with systematic differences in the informational quality of what audiences receive. This fact motivates an epistemic perspective that conceptualizes \emph{ad delivery not only as the allocation of attention, but also as the allocation of interpretative resources} - the shared concepts and meaning-making tools whose gaps and distortions are central to hermeneutical injustice. This perspective is complementary to recent work on representational harms in algorithmic systems, which examines how models and outputs can depict, classify, or rank groups in harmful ways, for example through stereotyping or underrepresentation \cite{katzman2023taxonomizing,wang2025measuring,kong2026point,lalor2024should}. While representational harm frameworks focus on harmful portrayals or classifications, our focus is on how ad delivery systems may differentially distribute the interpretative resources needed to make experiences intelligible. In this sense, representational harms can arise even when the relevant concepts are already available, whereas the hermeneutical injustice lens highlights cases in which a system helps sustain or reproduce interpretative gaps.

\subsection{An Epistemic Perspective on Online Advertising}

Digital platforms intensify segmentation beyond pre-digital media choice and geography by enabling fine-grained profiling and rapid content optimization \cite{wind1978issues,williamson1978decoding,oswald2012marketing,coeckelbergh2022political,coeckelbergh2025ai}. However, in online advertising, feature-driven targeting and personalization do not only lead to reallocation of attention, but also reallocation of interpretative resources embedded in ad content \cite{mccracken1986culture,pollay2014distorted}. As a consequence, they can produce opinion clustering and epistemic polarization, saturating some audiences with narrow frames while excluding others \cite{ribeiro2018media,phillips2023high,cinus2022effect,hartmann2025systematic}. In addition, personalization can isolate similarly situated individuals, weakening the conditions for collective recognition of structural harms \cite{milano2025algorithmic}. Meanwhile, platform optimization - which inherently tends to privilege majority behavior and engagement-friendly content - can amplify dominant narratives and suppress contested voices \cite{stewart2022perfect,coeckelbergh2022political}, shaping which interpretations become socially salient and authoritative. As a consequence, even if relevant categories and framings may circulate, targeting and delivery can make them selectively visible - or systematically skewed - for particular audiences \cite{lambrecht2019algorithmic,imana2025auditing,andreou2018investigating,skurka2024repeated}. These pervasive harms in ad delivery can be seen as forms of \emph{hermeneutical injustice}, because they systematically shape who has access to the concepts and framings needed to interpret experiences, thereby reproducing group-structured gaps in interpretative resources. They yield a \emph{misdistribution}\footnote{Going beyond the conceptualization in \cite{fricker2007epistemic,fricker2017evolving}, which frames hermeneutical injustice as arising from structural deficits in collective interpretative responses.} \cite{coady2017epistemic} of concepts and, hence, epistemic resources across specific respondent groups.

Despite these recognized harms, hermeneutical injustice remain comparatively under-operationalized in algorithmic fairness \cite{coliva2025hysteria,hopster2024socially,milano2025algorithmic,pozzi2023automated,stewart2022perfect}. Dominant approaches instead emphasize distributive criteria such as parity in exposure or outcomes, highlighting the need not only to include these forms of injustice into debates on fairness in online ad delivery, but also to explicitly account for them within targeting and delivery operations themselves.

\section{Hermeneutical Harms: Formalization and Experimental Evidence}\label{sec:3}

The theoretical framework for ad delivery outlined in Section~\ref{sec:2} highlights a limitation of the fair ad-distribution literature, which largely focuses on exposure and outcomes but does not explicitly account for epistemic wrongs in ad delivery. In particular, targeted advertising can wrong subjects in their capacity as knowers by shaping what they can learn, recognize, and make sense of through patterned delivery and framing. The core idea from Fricker is that hermeneutical injustice arises when people lack adequate shared interpretative resources (such as concepts, labels, or frames) for making sense of significant aspects of their experience \cite{fricker2007epistemic}. The harm is not merely unequal access to information, but a disadvantage in making one's own experience intelligible because the social circulation of interpretative resources is incomplete or distorted.
To make this 
explicit, we distinguish two forms of epistemic harm:
\begin{itemize}
    \item \emph{Epistemic exclusion}: 
    Concepts are withheld from particular audiences through systematic under-exposure.
    \item \emph{Epistemic bombing}: Systematic over-exposure to narrow or sensational framings saturates attention. 
\end{itemize}
We then interpret these epistemic harms through the lens of \emph{hermeneutical injustice}. Our proposed framework frames hermeneutical injustice as a collective interpretative gap \cite{dotson2011tracking,dotson2012cautionary,medina2013epistemology}, that is, a gap between shared interpretative resources and what is needed to make own experiences intelligible. Building on this 
,  we treat ad delivery as a mechanism that allocates interpretative resources. Here, repeated or skewed exposure may affect how recipients process and evaluate informational content over time \cite{hassan2021effects,li2025does}, yielding two hermeneutical manifestations of the above harms:
\begin{itemize}
    \item \emph{Hermeneutical deprivation}: Relevant informational content fails to reach particular audiences, undermining their ability to interpret an experience. This can be exemplified by Alzheimer's prevention campaigns, whose ads often target an older audience \cite{perez2024comparison,heger2020raising}. However, based on this allocation and content choice, younger users - including those potentially affected by early-onset Alzheimer's - may receive limited exposure and fail to connect symptoms to available care pathways \cite{loi2022time}.
    \item  \emph{Hermeneutical distortion}: Repeated delivery of narrow or sensational framings can saturate or simplify concepts, misleading interpretation. This phenomenon is visible in short-form mental health advertising. For example, ADHD-oriented ads often present broad, engagement-optimized symptom checklists \cite{karasavva2025double}, repeated exposure to which can encourage overgeneralization or self-misclassification \cite{de2025exploring,schiros2025misinformation}.
\end{itemize}
We distinguish delivery-level mechanisms from their interpretative effects. Therefore, in our framework, \emph{epistemic exclusion} refers to identity-structured under-exposure to interpretative resources in ad delivery, whereas \emph{epistemic bombing} refers to repeated or saturating exposure to narrow or sensational framings. \emph{Hermeneutical deprivation} and \emph{hermeneutical distortion} refer to the corresponding harms that may arise when these patterns affect recipients' capacity for sense-making. We also distinguish \emph{epistemic exclusion} from \emph{informational injustice} \cite{bagwala2024informational}. Informational injustice concerns cases in which an informational asymmetry is used to disadvantage less-informed agents. Epistemic exclusion, by contrast, concerns the resulting exclusion from participation in knowledge practices. It becomes hermeneutically relevant when the withheld content matters for sense-making.

\subsection{Evidence of Hermeneutical Harms from Data: Analysis of AIDS Advertising}
We conduct an exploratory analysis of the \emph{AIDS Advertising Evaluation} surveys (1986-1987)\footnote{To the best of our knowledge, nobody has used the \emph{AIDS Advertising Evaluation} surveys for analyses of this kind.} to demonstrate the harms that the epistemic mechanisms introduced above may produce, and to ground the subsequent formalization of our proposed approach.
We consider a non-algorithmic advertising dataset as our baseline before extending the analysis to algorithmic settings in simulation (Section~\ref{sec:6}). To this end, we retain only surveys with consistent coverage of the variables used in our analysis, as detailed in Appendix \ref{app:a}. We use the offline survey data as a methodological baseline to show that the proposed hermeneutical mechanisms can arise outside platform-mediated online delivery. This means that we do not intend to treat this dataset as a direct proxy for online advertising, but rather as an empirical baseline for studying the epistemic mechanisms of interest. Although platform-mediated delivery builds on familiar advertising logics \cite{goldfarb2014different}, online systems differ in important ways, including scale, optimization dynamics, targeting granularity, repeated exposure, and attention capture. These differences may affect the magnitude and form of the harms we study. In the absence of a comparable online advertising dataset containing the measures required for our analysis, we use this dataset as an illustrative offline baseline to isolate the epistemic mechanisms.

\begin{figure}[t]
  \centering
  \begin{subfigure}[t]{0.49\linewidth}
    \centering
    \includegraphics[height=4.3cm,width=6cm]{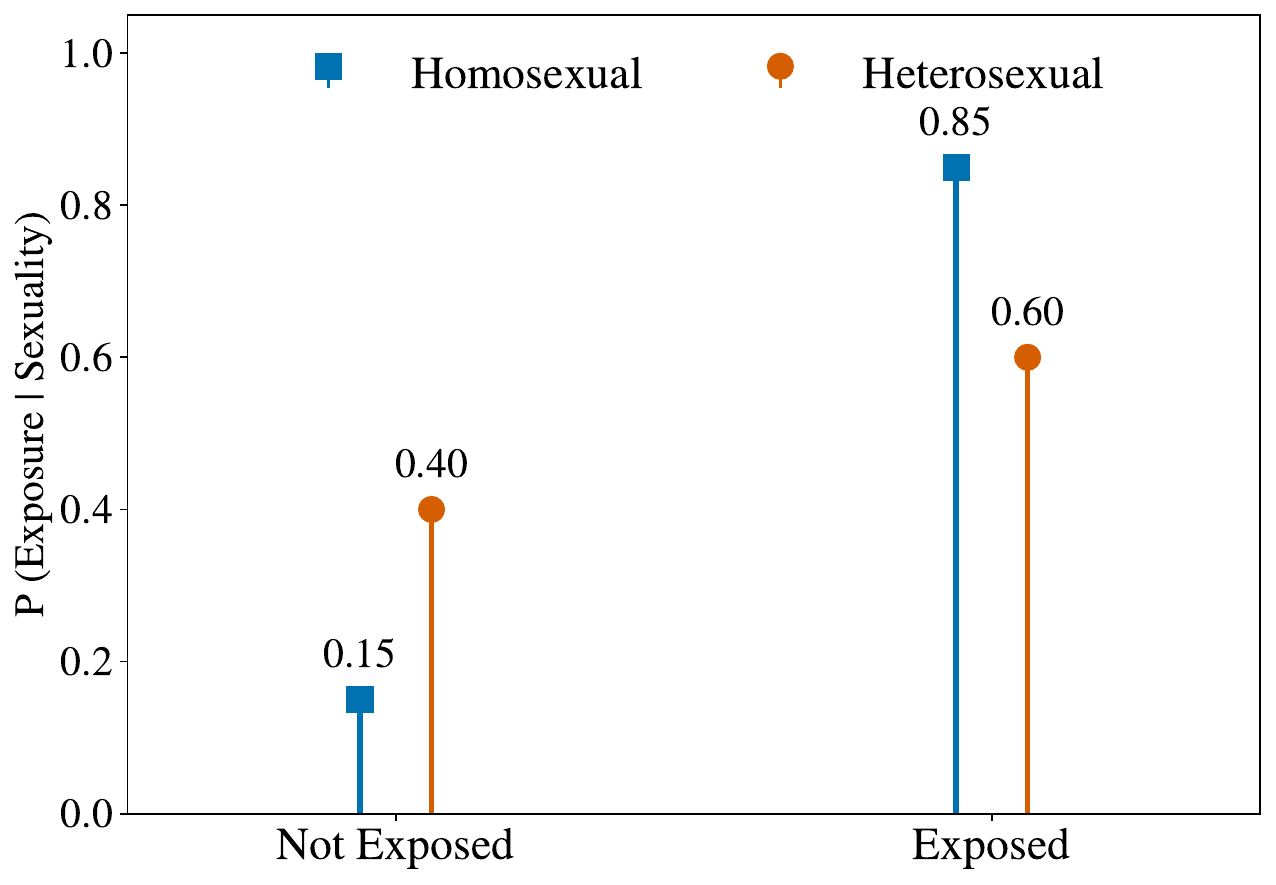}
    \caption{Exposure distributions $P(\text{Exposure}\mid\text{Sexuality})$.}
    \label{fig:bombing}
  \end{subfigure}\hfill
  \begin{subfigure}[t]{0.49\linewidth}
    \centering
    \includegraphics[width=\linewidth]{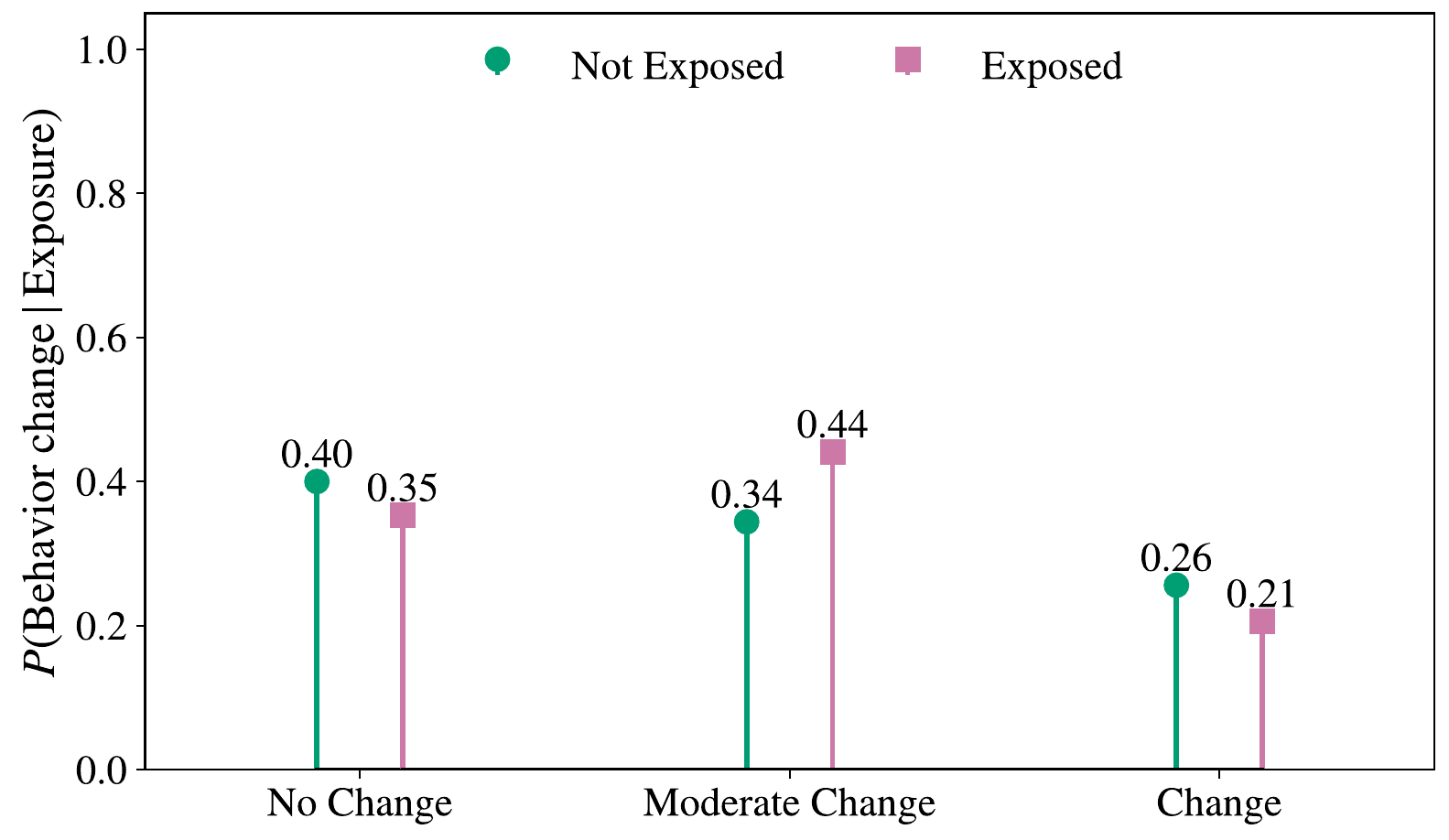}
    \caption{Behavior-change distributions $P(\text{Behavior change}\mid\text{Exposure})$.}
    \label{fig:uptake}
  \end{subfigure}
  \caption{\textbf{Asymmetric reach and non-monotonic behavior change in HIV prevention advertising.}
  (a) Asymmetric reach in HIV prevention advertising. Heterosexual respondents report a lower probability of high exposure and a higher probability of low exposure than homosexual respondents. 
  (b) Distribution of self-reported behavioral change after the campaign, stratified by exposure group (Not exposed vs Exposed; exposure categories defined in Appendix \ref{app:a}). Values report conditional proportions P(Behavior change|Exposure). In our data, higher exposure is not associated with a monotonic increase in substantial behavioral change: exposed respondents more often report moderate or no change than substantial change. 
  }
  \label{fig:bombing_uptake}
\end{figure}

Analyzing the available data, we observe two descriptive patterns that motivate our modeling of interpretative resource failures.\footnote{We do not treat the variables below as direct measurements of harm, but as illustrative indicators of harm-relevant patterns. In our reading, systematic under-exposure to relevant interpretative resources is consistent with deprivation, whereas exposure patterns that coincide with weak uptake or with responses suggestive of sensationalized or framing-sensitive reception are consistent with the possibility of distortion. A less harmful pattern would instead involve more even access to relevant exposure and stronger, less uneven uptake under exposure.} First, as shown in \figurename{~\ref{fig:bombing}}, exposure to ads is asymmetric. The results for heterosexuals report a systematically lower reach than for homosexuals, suggesting a clear targeting strategy in the advertising campaign and highlighting a potential unequal access to prevention-relevant concepts. Second, as shown in \figurename{~\ref{fig:uptake}}, self-reported behavioral change does not increase monotonically with exposure. When responses are stratified by exposure group, higher exposure is not associated with a larger share of substantial behavioral change. Instead, exposed respondents more often report moderate or no change than substantial change. We interpret this pattern as correlational evidence consistent with saturation-like dynamics, rather than as evidence of a monotonic increase in uptake with exposure. 

\begin{figure}[t]
\centering

\begin{subfigure}[t]{0.49\linewidth}
  \centering
  \includegraphics[width=\linewidth]{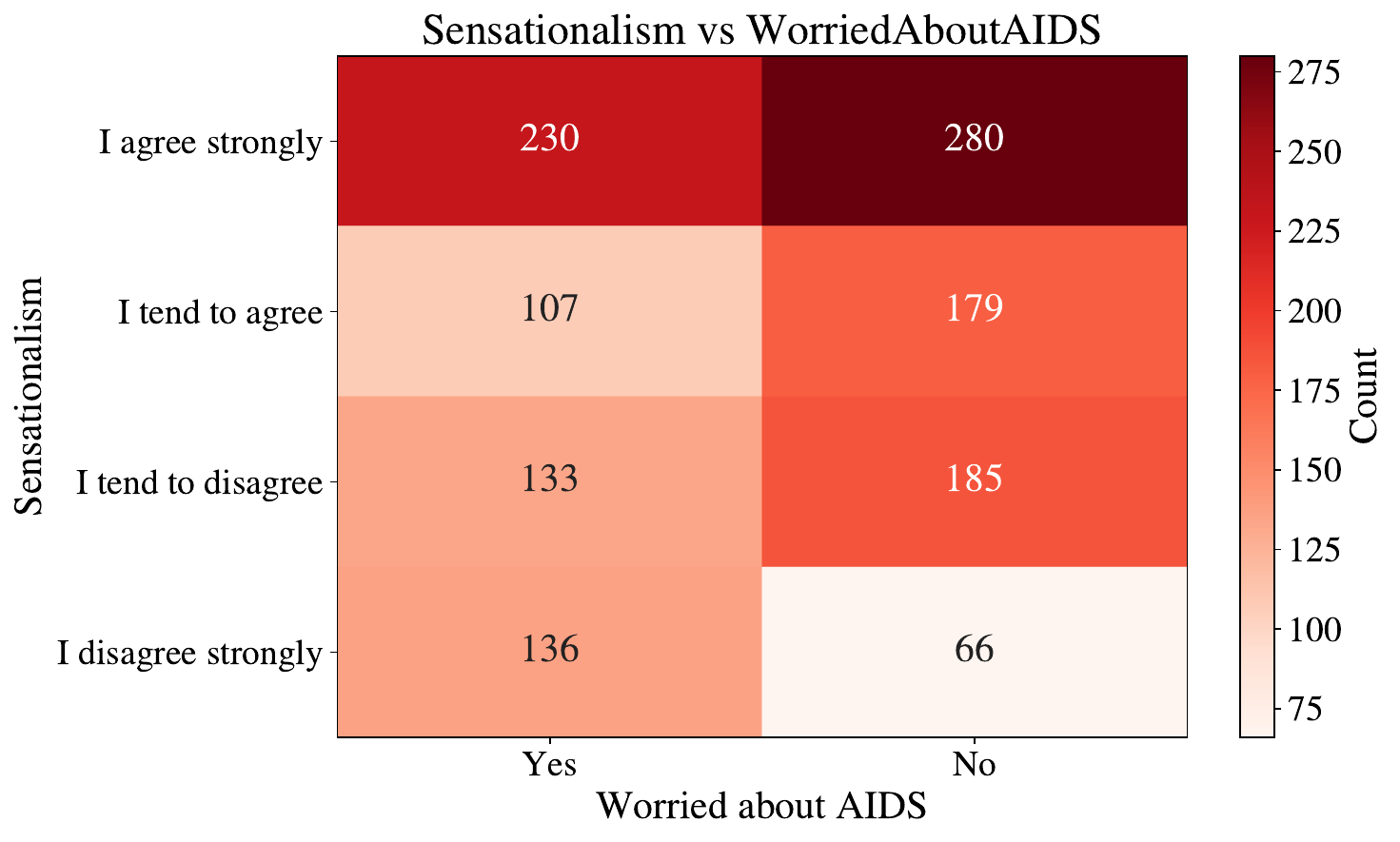}
  \caption{Sensationalism perceptions and worry about AIDS.}
  \label{fig:sensationalism_worry}
\end{subfigure}
\hfill
\begin{subfigure}[t]{0.49\linewidth}
  \centering
  \includegraphics[width=\linewidth]{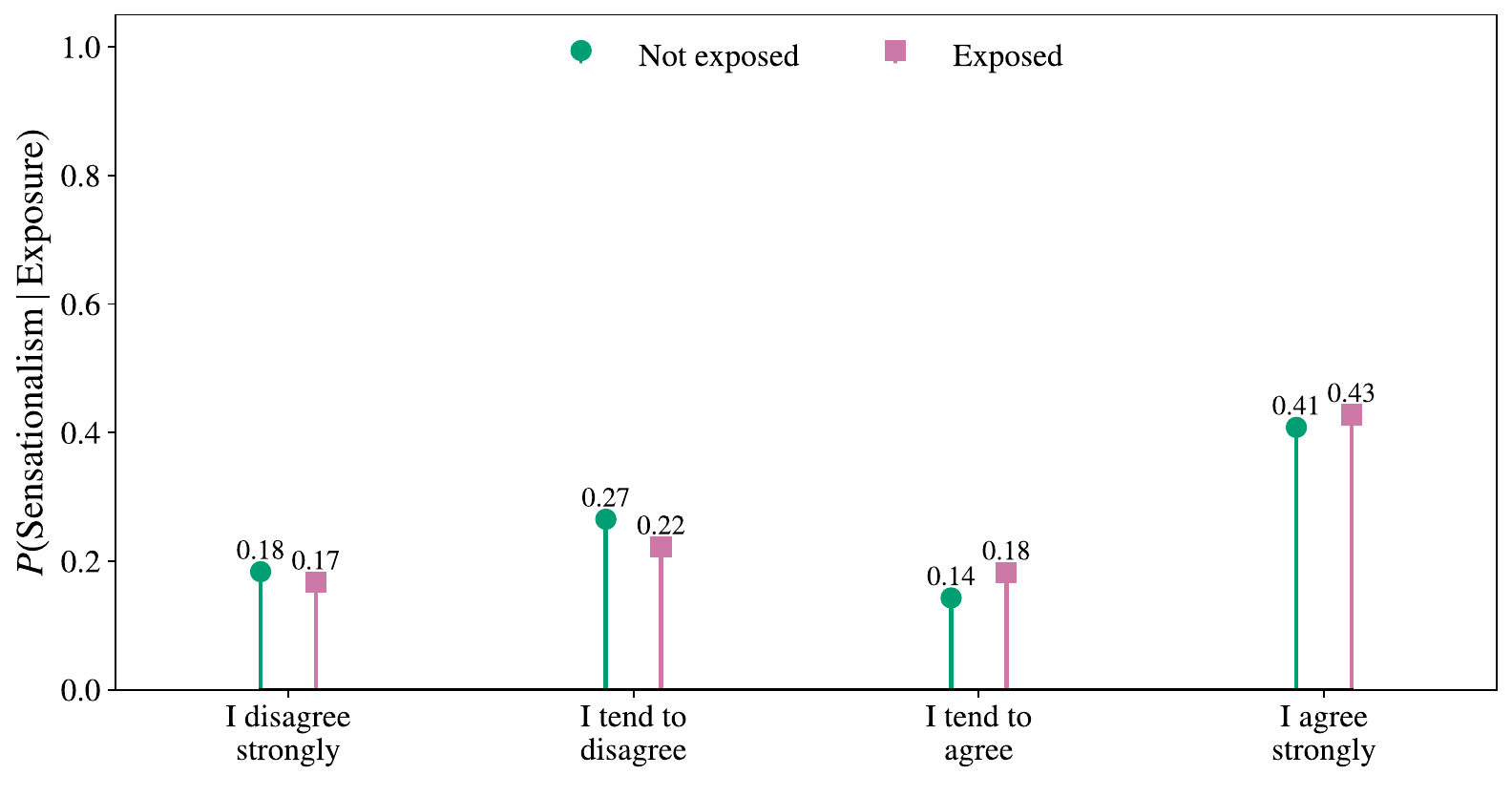}
  \caption{Sensationalism  $P(\text{Sensationalism}\mid\text{Exposure})$.}
  \label{fig:sensationalism_exposure}
\end{subfigure}

\vspace{4mm}

\begin{subfigure}[t]{\linewidth}
  \centering
  \includegraphics[width=0.49\linewidth]{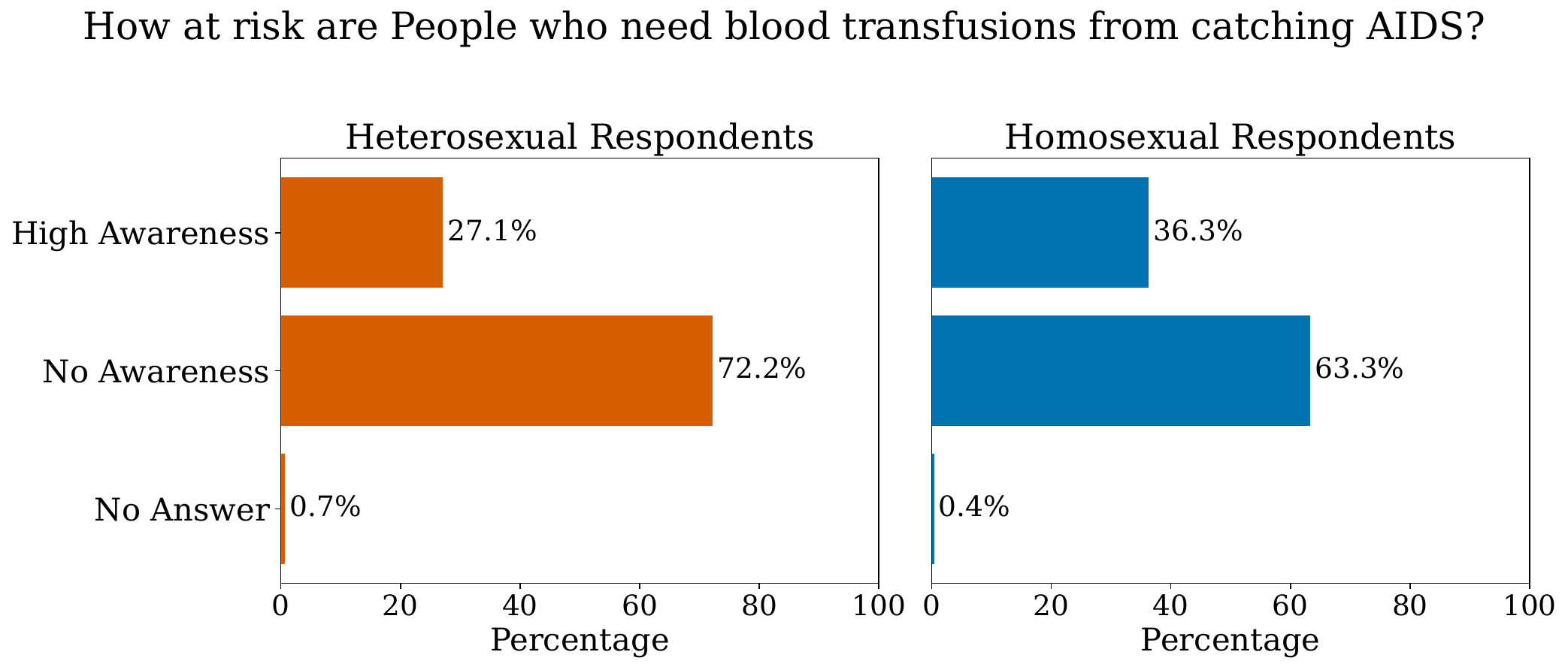}
  \hfill
  \includegraphics[width=0.49\linewidth]{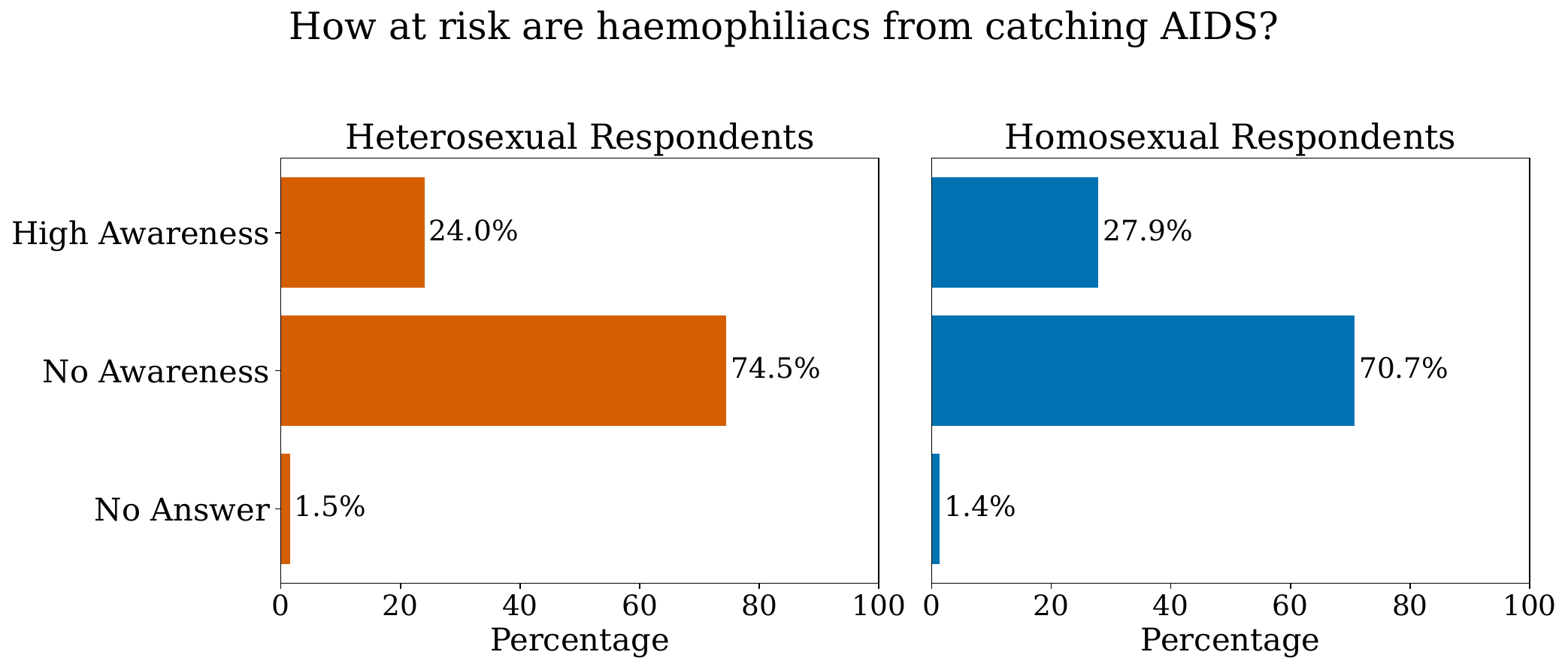}
  \caption{Framing-sensitive awareness of transfusion-related HIV risk.}
  \label{fig:transfusion_awareness}
\end{subfigure}

\caption{\textbf{Exploratory correlational evidence for deprivation and distortion mechanisms.}
(a) Perceived sensationalism vs. worry about AIDS (joint distribution). Count-annotated heatmap of the joint distribution between responses to the sensationalism item and self-reported worry about AIDS (Yes/No). The sensationalism item is: “How much do you personally agree with this statement?: The newspapers have been too sensational in their reporting of AIDS” (V154). Rows report responses (excluding Don’t know): I disagree strongly, I tend to disagree, I tend to agree, I agree strongly. 
(b) Sensationalism responses by exposure group. Conditional response distribution P(Sensationalism|Exposure) shown across categories (I disagree strongly, I tend to disagree, I tend to agree, I agree strongly) for Not exposed vs Exposed respondents.
(c) Transfusion-risk awareness variables exhibit framing sensitivity and lower self-reported awareness among heterosexual respondents.}
\label{fig:empirical_evidence}
\end{figure}

From the analyzed dataset, we also observe descriptive patterns consistent with the possibility of \emph{epistemic bombing} and related \emph{hermeneutical distortion}, while remaining agnostic about the underlying causal mechanism. As shown in \figurename{~\ref{fig:sensationalism_worry}}, stronger agreement with the statement that media coverage was ``too sensational'' is associated with lower self-reported worry about AIDS. In \figurename{~\ref{fig:sensationalism_exposure}}, the distribution of responses to the sensationalism item differs across exposure groups, with exposed respondents being relatively more likely to fall in the agreement categories than non-exposed respondents. Taken together, these patterns are consistent with the possibility that repeated or affectively skewed messaging may coincide with reduced perceived risk or weaker uptake of prevention messages. At the same time, these data are correlational and admit alternative explanations, including the possibility that respondents who were already less worried about AIDS were also more likely to perceive public discourse and advertising about AIDS as sensationalized. We wish to remark that our aim is not to establish a causal mechanism, but to show that delivery-uptake mismatches of the kind motivating hermeneutical concerns are visible in empirical data. 

The results of our analysis further indicate that interpretative resources are sensitive to framing, not only to exposure. Indeed, as shown in \figurename{~\ref{fig:transfusion_awareness}} for transfusion-related HIV risks, the subgroup-specific formulation (``people who need blood transfusions'' versus ``hemophiliacs'') yields response distributions that are comparable across groups of respondents. Both heterosexual and homosexual respondents are indeed more likely to report ``no awareness'' and less likely to report ``high awareness'', a pattern consistent with \emph{epistemic exclusion} induced not by exposure alone but by content framing. In turn, this indicates a potential group-free \emph{hermeneutical deprivation}, that is, reduced uptake of a prevention-relevant concept that can impede recognition and articulation of a risk. Appendix \ref{app:a} reports confidence intervals and basic association tests for the main descriptive patterns discussed.

Taken together, these associations provide exploratory evidence consistent with 
the two ways in which advertising can fail as a carrier of collective interpretative resources introduced at the beginning of this section
. Consequently, we draw on these empirically observed patterns as qualitative guidance to develop our formulation of hermeneutical injustice in ad delivery, which we then implement in the simulation study in Section \ref{sec:6}. Therefore, we treat these empirical patterns as qualitative, harm-relevant indicators that motivate the formalization developed in the remainder of the paper, rather than as definitive measurements of hermeneutical harm.


\section{Modeling Hermeneutical Fairness in Ad Delivery}\label{sec:4}
So far, we have argued that fairness in advertising has both distributive and hermeneutical dimensions, but the latter has not yet been formalized in the literature. In this section, we thus make this formalization step, building on the empirical evidence from Section~\ref{sec:3} and the utility-based framework from \cite{baumann2024fairness}. To this end, we focus on the (simplified) case in which the online ad platform makes a binary display decision $d_x \in \{0,1\}$ for each user $x$ on a focal ad. We further assume that the potential recipients of the ad form a finite set of users $G$, that can be partitioned into two protected groups $G_a$ and $G_b$ according to one or more features that influence the utility gained by the platform if they are targeted and their ability to understand or benefit from the ads  content. 
If the platform decides not to expose the user $x$ to the ad ($d_x=0$),  user $x$ can then be subject to \emph{hermeneutical deprivation}. This circumstance exacerbates \emph{selective underexposure} when individuals from a specific group are systematically excluded from receiving the ad. Such a selective advertising campaign can, at the same time, lead to \emph{epistemic bombing} among the targeted user groups (as evidenced in Section~\ref{sec:3}), due to \emph{systematic overexposure}. 

These issues can nonetheless be mitigated on the platform by designing ad delivery policies constrained to \emph{parity of exposure}, defined next.
\begin{definition}[Parity of exposure]
    Let $N_a$ and $N_b$ be the sizes of the two protected groups $G_a$ and $G_b$. We say the ad allocation decision $\{d_x\}_{x \in G}$ guarantees parity of exposure if the following holds
    \begin{equation}\label{eq:statistical_parity_exposure}
        \frac{1}{N_a}\sum_{x \in G_a} d_x=\frac{1}{N_b}\sum_{x \in G_b} d_x,
    \end{equation}
namely, if the average of ad receivers is equal for both groups.
\end{definition}
Note that this definition is nothing but the sample approximation for statistical parity of exposure (see the definition in \cite{corbett2017algorithmic}). Hence, hermeneutical underexposure and overexposure can be alleviated by enforcing classical distributive fairness constraints in ad-delivery policy design, thereby reinforcing our vision of hermeneutical fairness as an (interpretative) resource allocation problem. At the same time, underexposure and overexposure are not the only issues when looking at ad delivery from an epistemic perspective. Indeed, even if users receive an ad, they might still not make sense of it. Therefore, to achieve hermeneutical fairness, an ad platform must not only guarantee that both groups are reached by the advertisement but also ensure that different user groups can equally make sense of the ad content, according to the criteria formalized next.
\begin{definition}[Equality of hermeneutical opportunity]\label{def:equality_opportunity}
    Let $G_a$ and $G_b$ be two protected groups. Then, the ad allocation decision $\{d_x\}_{x \in G}$ guarantees equality of hermeneutical opportunity if the following holds
    \begin{equation}\label{eq:equality_opportunity}
        \frac{\sum_{x \in G_a} d_x\rho_x}{\sum_{x \in G_a} \rho_x}=\frac{\sum_{x \in G_b} d_x\rho_x}{\sum_{x \in G_b} \rho_x},
    \end{equation}
    with $\rho_x \in [0,1]$ being the probability that user $x$ understands the focal ad\footnote{With $\rho_x \approx 1$ indicating that user $x$ has successfully made sense of the ad's content.}.
\end{definition}
Note that, similarly to the definition employed in \cite{baumann2024fairness}, each side of \eqref{eq:equality_opportunity} corresponds to the share of users 
belonging to a specific user group seeing the focal ad 
among all those that would be able to make sense of it. 
\begin{remark}
   Based on our definition, equality of hermeneutical opportunity depends not only on the ad platform's delivery strategy, but also on the ad content. Hence, guaranteeing equality of hermeneutical opportunity implies redesigning ad campaigns that ultimately count as a cost for the platform.
\end{remark}
We can thus formally characterize a hermeneutically fair ad delivery strategy as follows:
\begin{definition}[Hermeneutically fair ad delivery]
    Consider two protected groups $G_a$ and $G_b$. We say that the ad allocation decision $\{d_x\}_{x \in G}$ is \emph{hermeneutically fair} if it guarantees parity of exposure as in \eqref{eq:statistical_parity_exposure} and equality of hermeneutical opportunity according to \eqref{eq:equality_opportunity}.  
\end{definition}

As already mentioned, allocating resources, both distributive and interpretative, has a cost. This motivates the adoption of a cost-sensitive perspective on hermeneutical fairness, which we will use in Section~\ref{sec:5} to inform the design of our hermeneutically-aware ad delivery strategy. Similarly to \cite{baumann2024fairness}, we can thus define a hermeneutic cost that depends on whether user $x$ receives the ad and uptake (i.e., makes sense of) the ad's content. Let $\rho_x \in [0,1]$ be the probability that user $x$ makes sense of the focal ad, where $\theta>0$ denotes the reward for successful uptake, $\omega>0$ the penalty for failed uptake, and $\xi>0$ the exclusion penalty for withholding the ad and its interpretative resources. We then
define the \emph{hermeneutical injustice cost} associated to user $x$ as
\begin{equation}\label{eq:hermenutical_injustice}
    \mathrm{HI}(d_x)= [-\theta \rho_x+\omega (1-\rho_x)]d_x+\xi(1-d_x).
\end{equation}
Hence, if the user does not receive the advertisement ($d_x=0$), the cost reduces to the fixed exclusion penalty $\xi$. If instead the user receives the ad ($d_x=1$), hermeneutical injustice decreases with $\rho_x$ at rate $\theta$ and increases with failed uptake, weighted by $\omega$.

Intuitively, according to \eqref{eq:hermenutical_injustice}, the ad can either increase utility, framed as $U(d_x)-{\rm HI} (d_x)$, (when uptake succeeds) or decrease it (when uptake fails). When the ad is withheld, by contrast, the user has no opportunity to benefit from its interpretative content, which we represent as a fixed exclusion cost $\xi$. 
\begin{remark}
     Similarly to the assumptions made in \cite{baumann2024fairness}, we suppose that the probabilities $\rho_x$ (or an estimate thereof) are available to the ad platform for all $x \in G$
     . In practice, $\rho_x$ is a latent variable that would likely not be directly observable. In modern advertising systems, it would more plausibly be inferred from richer platform-side signals, such as historical interaction patterns, repeated engagement, post-click behavior, and learned user/content representations. The controlled protocol outlined in Section \ref{sec:discussion} provides one possible strategy for validating or calibrating whether such platform-side proxies meaningfully track interpretative uptake.
\end{remark}
\begin{remark}
     In empirical applications, the epistemic parameters $\theta$, $\omega$, and $\xi$ could be estimated by measuring experimentally induced variation in exposure and framing, leading to controlled changes in recognition, articulation, or downstream actions attributable to withholding versus showing the ad. To do so, one would require data from controlled ad delivery experiments, whose collection and analysis we leave for future work.
\end{remark}

\section{Optimal Ad Delivery Strategy Under Hermeneutical Lens}\label{sec:5}
Relying on the previous definition of hermeneutical fairness and cost, we now extend the fairness-constrained utility maximization problem introduced in \cite{baumann2024fairness} to explicitly account for epistemic harms.

To this end, let the platform’s economic utility for a user $x$ be 
\begin{equation*}\label{eq:utility}
    u(d_x)= \alpha p_xd_x+\beta(1-d_x), 
\end{equation*}
where $p_x \in [0,1]$ is the platform's estimate of the probability that user $x$ takes the advertiser's desired action (e.g., clicks), $\alpha$ is the utility/revenue gained conditional on that action (e.g., the advertiser's bid), and $\beta$ is the utility when the high-stakes ad is not shown (e.g., the expected utility of the best alternative ad or showing no ad). Accordingly, let us define the expected utility over the overall set of users $G$ as
\begin{equation*}
    U(d)=\sum_{x \in G} u(d_x),
\end{equation*}
where $d=\{d_x\}_{x\in G}$.

To account for the economic impact of epistemic harms exemplified by \eqref{eq:hermenutical_injustice}, we assume that a rational decision maker seeks for $d=\{d_x\}_{x\in G}$ that maximizes the hermeneutically-aware utility
\begin{equation}\label{eq:overall_loss}
    \mathrm{U}^{\mathrm{H}}(d):=\mathrm{U}(d)-\gamma \mathrm{HI}(d),
\end{equation}
with
\begin{equation*}
    \mathrm{HI}(d)=\sum_{x \in G} \{[-\theta \rho_x+\omega (1-\rho_x)]d_x+\xi (1-d_x)\},
\end{equation*}
hence seeking for a trade-off between economic and epistemic expected utilities governed by the decision maker via the tunable trade-off parameter $\gamma \geq 0$.

Following \cite{baumann2024fairness}, we further assume that any acceptable allocation $d$ must satisfy baseline distributive and, eventually, hermeneutical fairness constraints. From a distributive fairness perspective, we focus on the classical group fairness criteria of statistical parity and equality of opportunity~\cite{corbett2017algorithmic}. The first choice ultimately allows us to already contrast under- and over-exposure\footnote{See the connection between statistical parity and parity of exposure highlighted in Section~\ref{sec:4}.}. Throughout the second, we impose
\begin{equation}\label{eq:equality_distributional}
        \frac{\sum_{x \in G_a} d_x p_x}{\sum_{x \in G_a} p_x}=\frac{\sum_{x \in G_b} d_x p_x}{\sum_{x \in G_b} p_x},
\end{equation}
where $p_x \in [0,1]$ is the probability of user $x$ clicking on the focal ad. 
In addition, we enforce \eqref{eq:equality_opportunity} as  a hermeneutical constraint, hence promoting user groups to equally make sense of the focal ad when received, leading to the \emph{hermeneutically-aware ad optimization problem} as
\begin{equation}\label{eq:maximization}
\begin{aligned}
    \max_d &\quad\mathrm{U}^{\mathrm{H}}(d)\\
    \mbox{s.t.} & \quad \eqref{eq:statistical_parity_exposure}, \eqref{eq:equality_opportunity}, \eqref{eq:equality_distributional}. 
\end{aligned}
\end{equation}
Accordingly, we can define the optimal allocation rule $d^{\star}(\gamma)$ as
\begin{equation}\label{eq:optimal_move}
\begin{aligned}
    & d^{\star}(\gamma)=\arg\max_d  \quad\mathrm{U}^{\mathrm{H}}(d)\\
    & \qquad \qquad \qquad \mbox{s.t.} \quad \eqref{eq:statistical_parity_exposure}, \eqref{eq:equality_opportunity}, \eqref{eq:equality_distributional},
\end{aligned}
\end{equation}
and the hermeneutically fair allocation rule as follows.
\begin{definition}[Hermeneutically Fair Allocation Rule]\label{def:HF_rule}
    The hermeneutically fair allocation rule is the mapping that returns $d^{\star}(\gamma)$ in \eqref{eq:optimal_move} when $\gamma>0$. 
\end{definition}
Therefore, in the absence of hermeneutically aware components in \eqref{eq:maximization}, a platform may allocate low-uptake impressions unevenly across groups. This is not the case in our formulation, where hermeneutical constraints and loss terms regulate the distribution of hermeneutical mechanisms, namely content deprivation and distortion, rather than merely the distribution of impressions.

\paragraph{Unconstrained closed-form solution} Since the objective is additive over $x$, in the absence of any distributive and/or hermeneutical constraints, the maximization of $\mathrm{U}^{\mathrm{H}}(d)$ in \eqref{eq:overall_loss} admits an explicit point-wise optimal solution. The associated optimizer is determined by showing the focal ad ($d_x=1$) if and only if
\begin{equation}\label{eq:ineq:1}
[u(d_x)-\gamma \mathrm{HI}(d_x)]d_x \geq [u(d_x)-\gamma \mathrm{HI}(d_x)](1-d_x),
\end{equation}
i.e., if the hermeneutically-aware utility is maximized if user $x$ receives the utility. Equivalently, \eqref{eq:ineq:1} corresponds to 
\begin{equation*}\label{eq:pointwise_rule}
\alpha p_x + \gamma \theta \rho_x-\gamma \omega(1-\rho_x)\geq\beta-\gamma\,\xi,
\end{equation*}
leading to the threshold that can be operationalized by the platform, namely
\begin{equation*}\label{eq:threshold}
p_x \geq \frac{\beta-\gamma\xi+\gamma\omega(1-\rho_x)-\gamma \theta\rho_x}{\alpha}.
\end{equation*}
This last rule highlights the resulting decision logic at the individual level in the absence of constraints. For fixed $\gamma$, the platform should prefer showing the focal ad to users with a higher uptake probability $\rho_x$. Given the combinatorial complexity of the active constraints given the parameters, no analogous closed-form characterization is provided for the constrained cases. Once fairness constraints are imposed, we solve the resulting optimization problem numerically for each parameter configuration in Section~\ref{sec:6}.

\section{Trade-offs Between Utility, Distributive Constraints and Hermeneutical Costs: A Simulation Study}\label{sec:6}

\begin{table}[!tb]
    \centering
    \begin{tabular}{ccccccccc}
        & \multicolumn{8}{c}{\textbf{Parameters}}\\
        \cline{2-9}
        \textbf{Scenario} & $\alpha$ & $\beta_a$ & $\beta_b$ &  $\theta_a$ & $\theta_b$ & $\omega_a$ & $\omega_b$ & $\xi$\\
        \hline
        \textbf{A} & 0.2 & 0.03 & \textbf{var.} & 0.05 & 0.1 & 0.01 & 0.01 & 0.2\\
        \hline
        \textbf{B} & 0.2 & 0.03 & 0.05 & 0.05 & \textbf{var.} & 0.01 & 0.01 & 0.2\\
        \hline 
        \textbf{C} & 0.2 & 0.03 & 0.05 & 0.05 & 0.1 & 0.01 & \textbf{var.} & 0.2\\
        \hline
        \textbf{D} & 0.2 & 0.03 & 0.05 & 0.05 & 0.1 & 0.01 & 0.01 & \textbf{var.}\\
        \hline
    \end{tabular}
    \caption{Numerical simulations: explored scenarios \emph{vs} fixed and varying (var.) utility and epistemic parameters.}\label{tab:parameters}
\end{table}

Through numerical simulations, we now analyze how variations in the trade-off weighting $\gamma$, as well as the utility and epistemic parameters in \eqref{eq:overall_loss} change the allocation policy and, hence, the resulting trade-offs between utility, classical fairness constraints, and hermeneutical harms. In doing so, we always assume $\alpha=0.2$, hence implying that the gain for receiving the focal ad is the same regardless of which user group they belong to. Meanwhile, we study the trade-off by dividing users into two groups, $a$ and $b$, fixing $\theta_a=0.05$ and $\beta_a=0.03$, while letting the remaining user-specific utility and epistemic parameters vary as reported in \tablename{~\ref{tab:parameters}}.

Across simulations, we compare five allocation rules: \emph{unconstrained}, where the platform only aims at maximizing the hermeneutical-aware utility $\mathrm{U}^{\mathrm{H}}(d)$; under \emph{parity of exposure}, i.e., when hermeneutical-aware utility is maximized subject to \eqref{eq:statistical_parity_exposure}; under \emph{equality of opportunity}, i.e., when hermeneutical-aware utility is maximized subject to \eqref{eq:equality_distributional}; with \emph{equality of hermeneutical opportunity}, so that hermeneutical-aware utility is maximized under \eqref{eq:equality_opportunity}. In all scenarios, $\gamma$ in \eqref{eq:overall_loss} is set to $0.01$. 
In each simulation run, we 
sample group-specific uptake probabilities $\rho_x$ from $\mathrm{Beta}$ distributions
. Concretely, we use $\rho_a \sim \mathrm{Beta}(4,6)$ and $\rho_b \sim \mathrm{Beta}(7,3)$. This yields mean uptake probabilities of $0.4$ and $0.7$, respectively, thus modeling group $b$ as having higher expected uptake than group $a$. We also consider three additional uptake configurations: an $a$-advantaged setting, with $\rho_a \sim \mathrm{Beta}(8,2)$ and $\rho_b \sim \mathrm{Beta}(3,7)$; a balanced high-uptake setting, with $\rho_a=\rho_b \sim \mathrm{Beta}(7,3)$; and a balanced lower-uptake setting, with $\rho_a=\rho_b \sim \mathrm{Beta}(4,6)$. The main text reports results for the $b$-advantaged case only. The corresponding results for the other uptake configurations are reported in Appendix \ref{app:b_1}.
We further model $p$ via a power-law distribution with fixed coefficients $k_a=0.05$ and $k_b=0.05$. For each scenario and each value of the varying parameters, we run 100 replications, resampling $\rho_x$ and $p$ in each of them.

Given this setting, the simulated scenarios are the following:
\begin{itemize}
    \item \textbf{A. Competitive spillover:} ($\beta_b$ varies, $\beta_a$ is fixed). This scenario models economic asymmetries in the opportunity cost of showing the focal ad. By varying the utility of the alternative ad for group $b$ ($\beta_b$) while keeping $\beta_a$ fixed, we simulate a market where one group is more expensive to target. This allows us to examine how the platform's allocation reacts to purely economic pressure.
    \item \textbf{B. Competitive uptake:} ($\theta_b$ varies, $\theta_a$ is fixed). In this case, we investigate how the model handles asymmetric benefits of understanding. By varying the reward for successful uptake $\theta_b$, we represent cases where the interpretative resource provided by the ad is more vital for one group than the other. The setup explores the tension between maximizing group-specific hermeneutical benefits and satisfying equality of opportunity constraints.
    \item \textbf{C. Competitive hermeneutical loss:} ($\omega_b$ varies, $\omega_a$ is fixed). This scenario isolates the risks associated with hermeneutical distortion. By varying the penalty $\omega_b$ for failed uptake in group $b$, we simulate high-stakes environments where epistemic bombing is particularly harmful to a specific subgroup. We use this to test how the platform balances the risk of epistemic harms against the requirement for parity of exposure.
    \item \textbf{D. Hermeneutical spillover:} (the exclusion cost $\xi$ varies). This parameter represents the cost of being systematically blind to the interpretative resources contained in the ad. This setup allows us to observe how the platform's preference for broad reach evolves as the cost of withholding information increases.
\end{itemize}

\section{Discussion}\label{sec:discussion}

\begin{figure}[t]
    \centering
    \includegraphics[scale=.30]{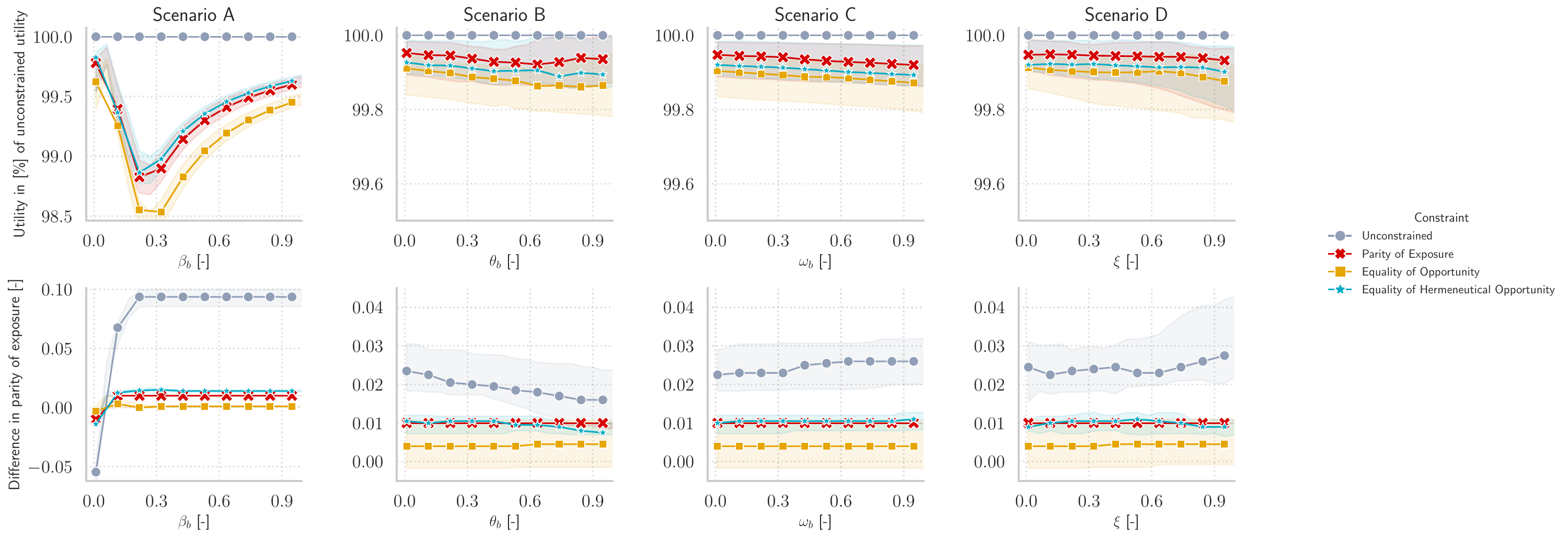}
    \caption{\textbf{Simulation results for Scenarios A--D (Table~\ref{tab:parameters})} imposing different fairness constraints. Top row: achieved utility (in \% of the unconstrained solution). Bottom row: parity of exposure difference. Solid lines show the median outcomes across the 100 replications, while the shaded bands show their range according to the 0.25-0.75 quartiles.}
    \label{fig:scenarios_paper}
\end{figure}

We discuss the results shown in Figure~\ref{fig:scenarios_paper} and the limitations of our work.
\subsection{The Price of Hermeneutical Fairness}
As shown in \figurename{~\ref{fig:scenarios_paper}}, imposing any of the considered fairness constraints reduces the platform's hermeneutically-aware utility, $U^{\rm H}(d)$, compared to the purely economic cost (baseline), $U(d)$. In particular, equality of opportunity induces the largest utility loss across all scenarios, while parity of exposure and equality of hermeneutical opportunity remain closer to the unconstrained utility. The magnitude of the price of hermeneutical fairness is modest. Indeed, even in the scenario with the largest drop in utility (A at low $\beta_b$), enforcing equality of opportunity reduces utility by at most a couple of percentage points. Meanwhile, the other two constraints lead overall to slightly smaller losses. Focusing on Scenario A, all constrained solutions move closer to the unconstrained optimum as $\beta_b$ increases. Meanwhile, in Scenarios B-D, utility under all constraints decreases slightly as the respective parameter increases, but the losses remain below a few percentage points, and the relative ordering of the constraints remains stable.

\subsection{Economic Pressure and Hermeneutical Deprivation}
The lower panels of \figurename{~\ref{fig:scenarios_paper}} highlight the distributional effects across the explored scenarios. Focusing on the unconstrained policy, the difference in parity of exposure is consistently positive (see the gray-circled curve in the bottom row of all panels), exceeding 10\% in some scenarios (most notably Scenario~A). This result implies that, by using the unconstrained delivery policy, the platform systematically under-delivers ads to the disadvantaged group. By contrast, imposing either of the three classical fairness constraints reduces these disparities to values close to zero (see the three curves in the bottom row corresponding to constrained scenarios in \figurename{~\ref{fig:scenarios_paper}}), thereby preventing the disadvantaged group from being structurally sidelined in ad delivery. From the platform's perspective, however, such corrections are not costless. Even though the efficiency losses we observe are numerically modest (on the order of 0.1–2\% relative to the unconstrained utility), they arise because the (hermeneutically-aware) platform is no longer free to choose an economic utility-maximizing policy. This claim is supported by the results in shown in the top row of \figurename{~\ref{fig:scenarios_paper}}, where curves corresponding to constrained scenarios consistently lie below the unconstrained benchmark. This fact creates economic pressure to ignore hermeneutical stakes. Importantly, the unconstrained policy still optimizes the hermeneutically-aware utility objective but does not impose a distributive constraint. In the bottom row of \figurename{~\ref{fig:scenarios_paper}}, the unconstrained solution exhibits a positive parity of exposure difference that becomes large under economic asymmetries (Scenario~A, as $\beta_b$ increases) and remains positive under group-$b$ specific penalties for failed uptake (Scenario~C, as $\omega_b$ varies). Our simulations thus illustrate a form of hermeneutical deprivation that is endogenous to the ad-delivery objective: without explicit distributive constraints, the optimizer may still withhold interpretatively valuable ads from the group for which delivery becomes comparatively less attractive under the modeled asymmetries. In particular, when we model group-$b$ specific penalties for failed uptake via $\omega_b$, the unconstrained policy maintains a positive exposure gap in Scenario~C (bottom-row gray curve in \figurename{~\ref{fig:scenarios_paper}}), indicating that the group facing higher modeled harm from failed uptake is also the group from which exposure is withheld.

\subsection{Hermeneutical Fairness Beyond Parity of Exposure}
Taken together, our results refine existing understandings of fairness-utility trade-offs in online ad delivery by shifting attention from exposure alone to its hermeneutical consequences. Our simulations suggest that once hermeneutical considerations are made explicit in the objective, the additional price of enforcing group-level fairness remains limited. Indeed, across Scenarios~A-D, the constrained policies achieve roughly 98-100\% of the unconstrained utility, with the largest observed gap being about 2\% in Scenario~A. At the same time, the same constraints substantially reduce the concentration of exposure disparities: the unconstrained solution exhibits a consistently positive parity of exposure difference across scenarios (gray curve in the bottom row plots of \figurename{~\ref{fig:scenarios_paper}}), reaching above 0.1 in Scenario~A, whereas the constrained variants remain near 0 throughout.
Within this setting, the classical fairness constraints we considered (i.e., statistical parity and equality of opportunity) thus play a dual role. On the one hand, they secure a more even distribution of impressions, as shown by the fact that the difference in parity of opportunity in all constrained cases is close to 0 (see the plots in the bottom panel of \figurename{~\ref{fig:scenarios_paper}}). At the same time, they help mitigate the systematic withholding of interpretative resources by preventing the large exposure gaps that arise under the unconstrained objective when asymmetries make delivery to one group less attractive (e.g., as shown in the unconstrained case in \figurename{~\ref{fig:scenarios_paper}} for Scenario~A as $\beta_b$ increases).
Meanwhile, our framework also highlights a crucial limitation of exposure-based fairness criteria. By construction, these constraints only regulate who sees a given ad, not how that ad interacts with existing hermeneutical asymmetries. Accordingly, policies that equalize exposure can still differ in their hermeneutical implications. This result is confirmed by comparing the red and blue-starred lines in \figurename{~\ref{fig:scenarios_paper}}, which highlight that imposing parity of exposure and equality of hermeneutical opportunity yield different utility levels, even though both produce similarly small parity-of-exposure differences. In principle, a policy can satisfy parity of exposure while still producing hermeneutical deprivation (for instance, by withholding key concepts from all groups) or hermeneutical distortion (for instance, by disproportionately saturating one group with low-uptake or skewed framings). This suggests that future fairness audits of ad-delivery systems should incorporate measures of interpretative opportunity and risk, rather than treating exposure parity as a sufficient proxy for fairness.

\subsection{Limitations} \label{subsec:limitations}
The main limitation of our 
framework concerns 
empirical validation. Our exploratory analysis relies on an offline advertising dataset that serves as a baseline rather than as a direct proxy for online ad delivery. This allows us to illustrate the epistemic mechanisms of interest, but it does not capture central features of online advertising such as large-scale optimization, fine-grained targeting, repeated delivery over time, and persistent competition for attention. These features may affect both the magnitude and the form of the hermeneutical harms we identify. The empirical results should therefore be read as illustrative
. A second limitation concerns parameter identification. In our model, the epistemic parameters are specified a priori rather than identified from data. In particular, the historical survey data used in our exploratory analysis do not contain a direct measure of uptake probability $\rho_x$. 
The available variables capture exposure, beliefs, awareness, perceived knowledge, confusion, and some downstream behavioral proxies. These 
features do not allow to isolate uptake itself from its downstream consequences. 

A concrete path toward empirical validation is a controlled ad-delivery study designed to estimate individual-level uptake probability $\rho_x$. Participants would be assigned to attribute-myopic delivery conditions that vary exposure intensity and framing, and would then complete a post-exposure uptake evaluation combining content understanding, scenario-based application, and a teach-back or paraphrase task \cite{baseman2013public,talevski2020teach}. The resulting responses could be aggregated into an individual uptake score yielding an estimate $\hat{\rho}_x$. Appendix \ref{app:c} reports the corresponding flowchart and provides additional details.
More broadly, an auditing workflow would require estimating and validating $\hat{\rho}_x$, calibrating epistemic parameters, evaluating group-level exposure and hermeneutical opportunity constraints where protected-group information is available, and stress-testing robustness to proxy misspecification and estimation error. A further failure mode concerns over-personalized interpretability. Optimizing for predicted uptake may improve average comprehension while reducing plural interpretability or worsening outcomes for specific subgroups.

\subsection{Ethical Considerations}\label{subsec:ethical}

We are aware that our framework might raise ethical concerns. At the same time, the level of analysis we develop in this paper remains primarily conceptual, rather than directly deployable in real-world ad systems. In the empirical analysis, the protected-group variable is a coarse proxy based on reported recent partner history, not a measure of sexual identity. Any grouping induced from these variables may misclassify respondents, collapse distinct experiences, and reproduce stigmatizing categories. These risks are further amplified by substantial missingness and uneven survey coverage.
The framework should likewise not be read as licensing intrusive profiling in real-world ad systems. While access to protected attributes and per-user uptake estimates is analytically useful for formalization and auditing, inferring such quantities in practice may create privacy risks, stigmatizing segmentation, and manipulative optimization. In particular, optimizing on predicted uptake may privilege commercially effective but overly narrow interpretative framings. For this reason, any empirical implementation of the framework would require strict data minimization, strong governance constraints on sensitive attributes, validation of uptake proxies, and auditing for failure modes including misclassification, proxy misspecification, over-personalized interpretability, and reduced plural interpretability.

\section{Conclusions and Future Work}\label{sec:conclusions}

Fairness in advertising is often cast as a purely distributive problem: ensuring that impressions or outcomes are comparable across protected groups. While this perspective has yielded important insights, our analysis indicates that it leaves out a crucial epistemic dimension. Building on Fricker's notion of hermeneutical injustice, we treat ads as carriers of interpretive resources and formalize two hermeneutical harms. On the one hand, we introduce hermeneutical deprivation through systematic under-exposure. On the other hand, we define hermeneutical distortion as a systematic over-exposure. We illustrate these mechanisms using exploratory correlational evidence from the \emph{AIDS Advertising Evaluation} surveys, where framing-sensitive awareness gaps and patterns involving perceived sensationalism are consistent with deprivation and distortion. We then incorporate these harms into an allocation framework by defining a new, hermeneutically aware utility and a group-level constraint on hermeneutical opportunity, thereby extending standard utility-driven and parity-based formulations. Controlled simulations trace how hermeneutical harms interact with classical fairness constraints, indicating that classical fairness requirements come with a modest but non-zero loss in hermeneutically-aware utility. Meanwhile, equality of opportunity is consistently the constraint that affects cost the most. At the same time, classical fairness constraints substantially reduce concentrated under-delivery to disadvantaged groups. This explicitly highlights that economic utility, distributive guarantees, and hermeneutical harms are not automatically aligned in ad delivery.

More broadly, this paper advances a broader aim of informing formal allocation models with philosophically grounded epistemic concepts. The framework shows how hermeneutical injustice can be embedded into optimization objectives and constraints, reframing fairness in ad delivery as  a hermeneutical question. Indeed, it is not only who sees which ads that matters, but also what interpretative resources those ads provide and how different groups are positioned to make sense of them. Incorporating hermeneutical harms into ad allocation provides a principled approach to detecting and mitigating exclusionary dynamics that may persist under exposure parity alone.

Future work includes: $i) $ estimating our model parameters and validating uptake-related constructs and our framework within controlled laboratory experiments and, where possible, using online ad-delivery data; $ii)$ validating our notion of hermeneutical fairness in other domains, 
such as job advertising, finance, and politics; $iii)$ extend our framework to multiple protected groups, multiple ads, and to account for dynamic feedback.

\bibliography{sample-base}
\bibliographystyle{ACM-Reference-Format}


\appendix

\section{AIDS Advertising Evaluation (1986-1987): Dataset Details and Pre-processing}\label{app:a}

We use the \emph{AIDS Advertising Evaluation} surveys collected in 1986-1987. The public release includes multiple survey versions (questionnaire variants) fielded across several waves. We construct a pooled respondent-level dataset from versions 1, 2, 3, 4, 6, 7, 8, and 16. We instead exclude versions 5 and 12-15 because they omit key variables required for our constructs, including sexual behavior history questions needed to construct the protected-group proxy and behavioral outcome index. In the following, we explain how we have pre-processed data and used them for our analysis.

\paragraph{Protected-group proxy construction}
The dataset does not contain a direct measure of sexual orientation. We therefore define a coarse partner-history grouping from reported partners in the last 12 months:

\begin{itemize}
    \item \emph{Same-sex partner history} (``homosexual'' proxy): \texttt{SameSexPartners} $> 0$ (regardless of \texttt{OppositeSexPartners}).
    \item \emph{Opposite-sex only} (``heterosexual'' proxy): \texttt{SameSexPartners} $= 0$ and \texttt{OppositeSexPartners} $> 0$.
    \item \emph{Unknown}: otherwise (i.e., \texttt{SameSexPartners} $=0$ and \texttt{OppositeSexPartners} $=0$, or missing partner counts).
\end{itemize}

Respondents reporting both partner types are assigned to the any same-sex group. This avoids forcing bisexual respondents into the opposite-sex-only group, but it collapses bisexuality into a broader non-heterosexual category. The proxy captures reported recent behavior (12 month window), not identity, and may misclassify respondents whose relevant behavior occurs outside the window or is not disclosed. We therefore interpret subgroup comparisons as descriptive and potentially conservative with respect to the any same-sex group.

Table \ref{tab:group_counts_by_version} reports counts in the proxy-defined sample. The dataset contains $n=3{,}199$ coded respondents, comprising $2{,}097$ classified as opposite-sex only and $1{,}102$ classified as any same-sex. As indicated in Table \ref{tab:group_counts_by_version}, the sample is unbalanced overall and exhibits pronounced imbalance across survey versions.

\begin{table}[t]
\centering
\small
\begin{tabular}{lrrr}
\textbf{Survey version} & \textbf{Opposite-sex only} & \textbf{Any same-sex} & \textbf{Total coded} \\
\midrule
version 1  & 654 &  40 & 694 \\
version 2  &   5 & 151 & 156 \\
version 3  & 660 &  40 & 700 \\
version 4  &   9 & 289 & 298 \\
version 6  & 216 &  10 & 226 \\
version 7  & 553 &  38 & 591 \\
version 8  &   0 & 284 & 284 \\
version 16 &   0 & 250 & 250 \\
\midrule
\textbf{Total} & \textbf{2097} & \textbf{1102} & \textbf{3199} \\
\bottomrule
\end{tabular}
\caption{Counts by survey version for the partner-history proxy, restricted to respondents for whom the proxy is defined.}
\label{tab:group_counts_by_version}
\end{table}

\paragraph{Advertising exposure measure}
We operationalize self-reported advertising exposure using reception variables of the form: ``Where did you see or hear this information?'' We treat each channel indicator as binary (Yes/No) where available (e.g., magazines, newspapers, radio, TV, pamphlets/leaflets, posters, newspaper articles, radio/TV programmes, talking with friends, telephone line).

We define respondent-level exposure as:

\begin{itemize}
    \item \emph{Exposed} if the respondent answers ``Yes'' to at least one observed reception variable.
    \item \emph{Not exposed} if the respondent answers ``No'' to all observed reception variables.
\end{itemize}

Exposure is coded missing if all reception variables are missing, to avoid treating nonresponse as non-exposure.

\paragraph{Behavioral-outcome proxy construction}
We construct a three-level behavioral-outcome proxy from two self-report binary variables:

\texttt{cut\_down\_sexual\_partners} and \texttt{used\_sheaths} (condoms).
We code:

\begin{center}
No Change (No/No), \quad Moderate Change (Yes/No or No/Yes), \quad Change (Yes/Yes).
\end{center}

This coding treats the two reported behaviors as complementary forms of risk reduction: fewer partners reduce exposure opportunities, while condom use reduces per-act transmission risk \cite{shelton2004partner,giannou2016condom,weller1996condom}.
We only define the composite index when both components are observed and standard (Yes/No) responses; otherwise the index is missing.

\paragraph{Missingness in behavioral variables and implications}
The behavioral variables exhibit substantial nonresponse in the dataset used for the behavioral-outcome analysis. Table \ref{tab:missingness_behavioral} reports missingness rates by partner-history group in this analysis sample.

To assess whether missingness is associated with the partner-history grouping, we test independence between a missingness indicator and partner-history group using Pearson $\chi^2$ tests on the corresponding contingency tables (Table \ref{tab:chi2_missingness}). In both cases, independence is rejected, indicating that missingness is systematically associated with the grouping. Accordingly, all statistics involving these behavioral variables are computed on the subset of non-missing responses and should be interpreted as descriptive patterns for that responding subset rather than as population-representative estimates.

\begin{table}[t]
\centering
\renewcommand{\arraystretch}{1.15}
\begin{tabular}{llrrrr}
\textbf{Variable} & \textbf{Partner-history group} & \textbf{Total} & \textbf{Observed} & \textbf{Missing} & \textbf{Missing rate (\%)} \\
\midrule
 {\small{\texttt{used\_sheaths}}} & Opposite-sex only & 2088 & 63  & 2025 & 96.98 \\
 \cline{2-6}
& Any same-sex      & 529  & 46  & 483  & 91.30 \\
 \cline{2-6}
& Unknown           & 169  & 0   & 169  & 100.00 \\
\hline
{\small{\texttt{cut\_down\_sexual\_partners}}} &
Opposite-sex only & 2088 & 63  & 2025 & 96.98 \\
 \cline{2-6}
& Any same-sex      & 529  & 233 & 296  & 55.95 \\
 \cline{2-6}
& Unknown           & 169  & 0   & 169  & 100.00 \\
\bottomrule
\end{tabular}
\caption{Missingness rates for the behavioral variables by partner-history group in the analysis sample used for the behavioral-outcome proxy.}
\label{tab:missingness_behavioral}
\end{table}

\begin{table}[t]
\centering
\renewcommand{\arraystretch}{1.15}

\begin{tabular}{lrrr}
{\small{\textbf{Variable}}} & $\boldsymbol{\chi^2}$ & \textbf{dof} & \textbf{$p$-value} \\
\midrule
{\small{\texttt{used\_sheaths}}}               & 43.527  & 2 & $3.5 \times 10^{-10}$ \\
\hline
\texttt{cut\_down\_sexual\_partners} & 769.586 & 2 & $7.7 \times 10^{-168}$ \\
\bottomrule
\end{tabular}
\caption{Results of Pearson $\chi^2$ tests of independence between missingness in the behavioral variables and partner-history group in the analysis sample, with degrees of freedom (dof) and $p$-value.}
\label{tab:chi2_missingness}
\end{table}

\paragraph{Statistical checks for the exploratory patterns}

For Figure \ref{fig:bombing}, we report 95\% Wilson confidence intervals for the conditional proportions $P(\text{Exposure}\mid\text{Sexuality})$, along with a Pearson $\chi^2$ test of independence for the associated $2 \times 2$ contingency table and Cram\'er's $V$ as a measure of association strength. Table \ref{tab:fig1a_ci} reports the estimated proportions. The association is statistically clear ($\chi^2(1)=148.37$, $p<0.001$, Cram\'er's $V=0.215$), indicating different exposure distributions across the two partner-history groups. Consistent with the framing of this section, we interpret this result as evidence of an asymmetric descriptive exposure pattern in the sample rather than as a definitive measure of harm.

\begin{table}[t]
\centering
\small
\begin{tabular}{llcccc}
\textbf{Group} & \textbf{Exposure} & \textbf{Count} & \textbf{N} & \textbf{Proportion} & \textbf{95\% CI} \\
\midrule
Any same-sex      & Not Exposed  & 883  & 1102 & 0.801 & [0.777, 0.824] \\
Any same-sex      & Exposed & 219  & 1102 & 0.199 & [0.176, 0.223] \\
Opposite-sex only & Not Exposed  & 1975 & 2097 & 0.942 & [0.931, 0.951] \\
Opposite-sex only & Exposed & 122  & 2097 & 0.058 & [0.049, 0.069] \\
\bottomrule
\end{tabular}
\caption{Estimated conditional proportions for Figure \ref{fig:bombing} with 95\% Wilson confidence intervals.}
\label{tab:fig1a_ci}
\end{table}

For Figure \ref{fig:uptake}, we report 95\% Wilson confidence intervals for the conditional proportions $P(\text{Behavior change}\mid \text{Exposure})$, together with a Pearson $\chi^2$ test of independence for the corresponding $2 \times 3$ contingency table and Cram\'er's $V$. Table \ref{tab:fig1b_ci} reports the estimated proportions. Confidence intervals are wider for the exposed group because of its smaller sample size. The global test does not indicate a statistically clear difference in the distribution of behavioral change across exposure groups ($\chi^2(2)=1.12$, $p=0.572$, Cram\'er's $V=0.084$). We therefore treat this pattern as descriptive and exploratory rather than as evidence of a strong difference between exposure groups.

\begin{table}[t]
\centering
\small
\begin{tabular}{llcccc}
\textbf{Exposure} & \textbf{Behavior change} & \textbf{Count} & \textbf{N} & \textbf{Proportion} & \textbf{95\% CI} \\
\midrule
Not Exposed  & No Change        & 50 & 125 & 0.400 & [0.318, 0.488] \\
Not Exposed  & Moderate Change  & 43 & 125 & 0.344 & [0.266, 0.431] \\
Not Exposed  & Change           & 32 & 125 & 0.256 & [0.188, 0.339] \\
Exposed & No Change        & 12 & 34  & 0.353 & [0.215, 0.521] \\
Exposed & Moderate Change  & 15 & 34  & 0.441 & [0.289, 0.605] \\
Exposed & Change           & 7  & 34  & 0.206 & [0.103, 0.368] \\
\bottomrule
\end{tabular}
\caption{Estimated conditional proportions for Figure \ref{fig:uptake} with 95\% Wilson confidence intervals.}
\label{tab:fig1b_ci}
\end{table}

For Figure \ref{fig:sensationalism_worry}, we report a Pearson $\chi^2$ test of independence for the corresponding $4 \times 2$ contingency table, together with Cram\'er's $V$. Table \ref{tab:fig2a_counts} gives the underlying counts. The association between sensationalism responses and self-reported worry about AIDS is statistically clear ($\chi^2(3)=47.87$, $p<0.001$, Cram\'er's $V=0.191$). As elsewhere in this appendix, we interpret this result as correlational support for the descriptive pattern shown in the heatmap, not as evidence of a causal effect.
\begin{table}[t]
\centering
\small
\begin{tabular}{lrr}
\textbf{Sensationalism} & \textbf{Yes} & \textbf{No} \\
\midrule
I disagree strongly & 136 & 66 \\
I tend to disagree  & 133 & 185 \\
I tend to agree     & 107 & 179 \\
I agree strongly    & 230 & 280 \\
\bottomrule
\end{tabular}
\caption{Contingency table for Figure \ref{fig:sensationalism_worry}, relating sensationalism responses to self-reported worry about AIDS.}
\label{tab:fig2a_counts}
\end{table}
For Figure \ref{fig:sensationalism_exposure}, we report 95\% Wilson confidence intervals for the conditional proportions $P(\text{Sensationalism}\mid\text{Exposure})$, together with a Pearson $\chi^2$ test of independence for the corresponding $2 \times 4$ contingency table and Cram\'er's $V$. Table \ref{tab:fig2b_ci} reports the estimated proportions. Confidence intervals are wider for the not-exposed group because of its smaller sample size. The global test does not indicate a statistically clear difference in the distribution of sensationalism responses across exposure groups ($\chi^2(3)=0.91$, $p=0.824$, Cram\'er's $V=0.032$). We therefore treat this pattern as descriptive and exploratory rather than as evidence of a strong between-group difference.
\begin{table}[t]
\centering
\small
\begin{tabular}{llcccc}
\textbf{Exposure} & \textbf{Sensationalism} & \textbf{Count} & \textbf{N} & \textbf{Proportion} & \textbf{95\% CI} \\
\midrule
Not exposed & I disagree strongly & 9   & 49  & 0.184 & [0.100, 0.314] \\
Not exposed & I tend to disagree  & 13  & 49  & 0.265 & [0.162, 0.403] \\
Not exposed & I tend to agree     & 7   & 49  & 0.143 & [0.071, 0.267] \\
Not exposed & I agree strongly    & 20  & 49  & 0.408 & [0.282, 0.548] \\
Exposed     & I disagree strongly & 144 & 860 & 0.167 & [0.144, 0.194] \\
Exposed     & I tend to disagree  & 191 & 860 & 0.222 & [0.196, 0.251] \\
Exposed     & I tend to agree     & 157 & 860 & 0.183 & [0.158, 0.210] \\
Exposed     & I agree strongly    & 368 & 860 & 0.428 & [0.395, 0.461] \\
\bottomrule
\end{tabular}
\caption{Estimated conditional proportions for Figure \ref{fig:sensationalism_exposure} with 95\% Wilson confidence intervals.}
\label{tab:fig2b_ci}
\end{table}
For Figure \ref{fig:transfusion_awareness}, we report Pearson $\chi^2$ tests of independence for the relevant contingency tables, together with Cram\'er's $V$. Under the formulation referring to \emph{people who need blood transfusions}, the distribution of awareness responses differs across the two partner-history groups ($\chi^2(2)=29.88$, $p<0.001$, Cram\'er's $V=0.097$). Under the \emph{haemophiliacs} formulation, the association is weaker and only borderline significant ($\chi^2(2)=5.99$, $p=0.050$, Cram\'er's $V=0.043$).
We also test whether response distributions differ across the two formulations within each partner-history group. The wording effect is statistically clear both among opposite-sex-only respondents ($\chi^2(2)=10.45$, $p=0.005$, Cram\'er's $V=0.050$) and among respondents with any same-sex partner history ($\chi^2(2)=22.94$, $p<0.001$, Cram\'er's $V=0.102$). As elsewhere in this section, we interpret these findings as correlational support for framing-sensitive awareness patterns rather than as definitive measures of harm.

\section{Additional Simulation Diagnostics}\label{app:sim_diagnostics}

\begin{figure*}[t]
\centering
\begin{subfigure}[t]{0.4\textwidth}
  \centering
  \includegraphics[width=\linewidth]{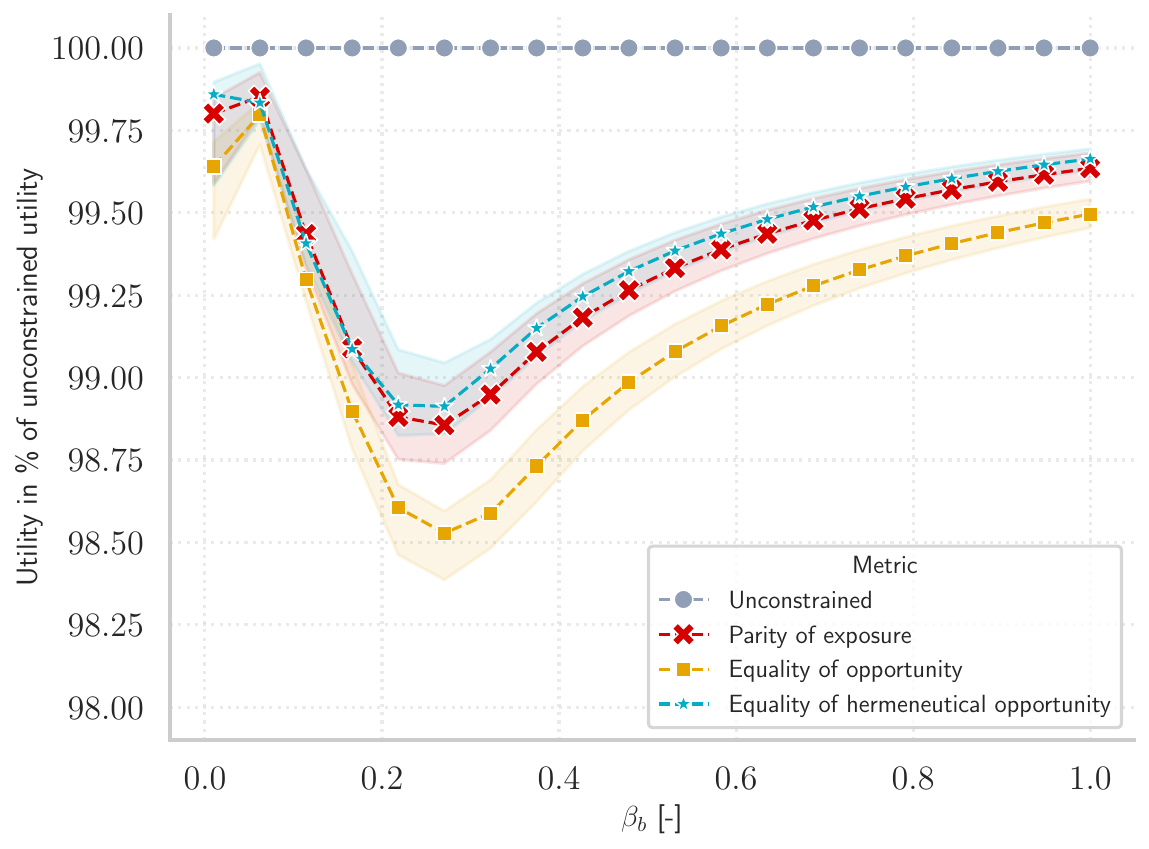}
  \caption{Utility (\% of unconstrained).}
  \label{fig:sim_beta_utility_app}
\end{subfigure}\hspace*{.75cm}
\begin{subfigure}[t]{0.4\textwidth}
  \centering
  \includegraphics[width=\linewidth]{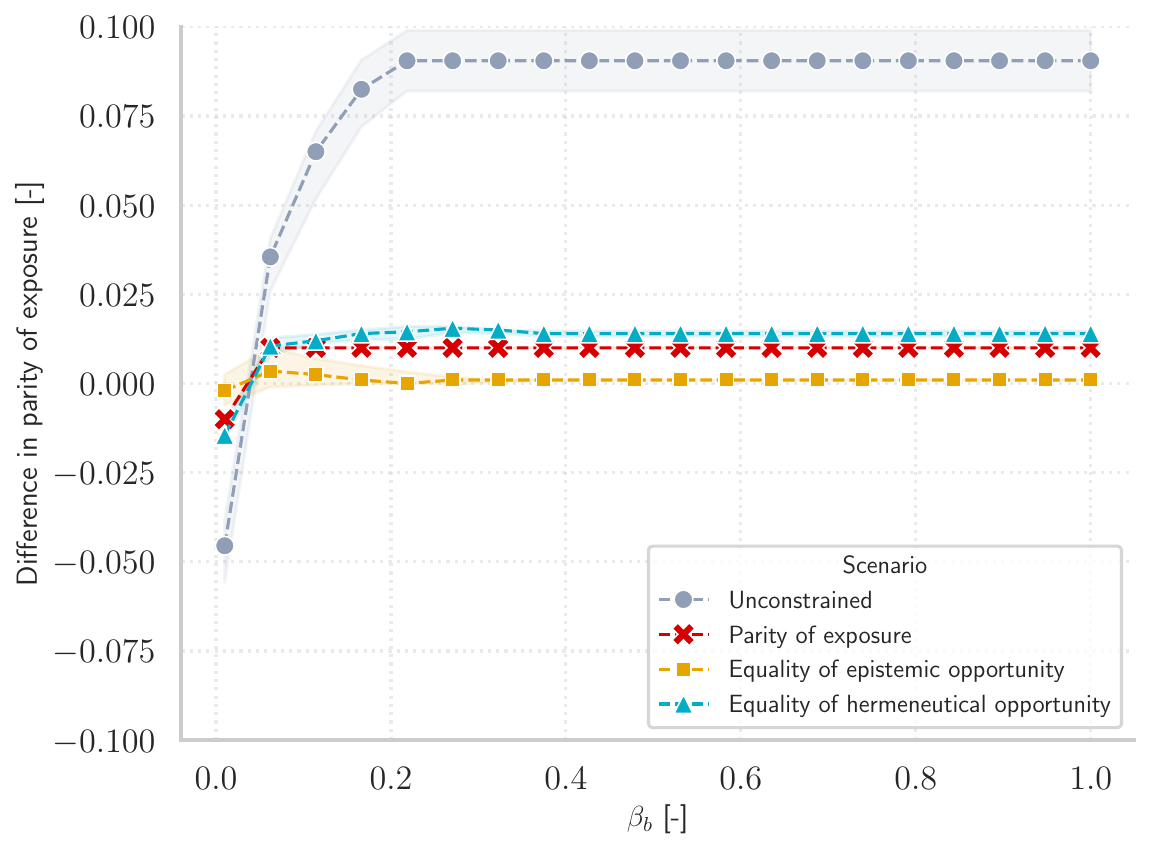}
  \caption{Parity of exposure difference.}
  \label{fig:sim_beta_parity_app}
\end{subfigure}
\caption{\textbf{Baseline results.} Achieved utility (as \% of the unconstrained solution) and parity of exposure difference for $\gamma=0$ and varying $\beta_b$.}
\label{fig:baseline}
\end{figure*}

\begin{figure*}[t]
\centering
\begin{subfigure}[t]{0.4\textwidth}
  \centering
  \includegraphics[width=\linewidth]{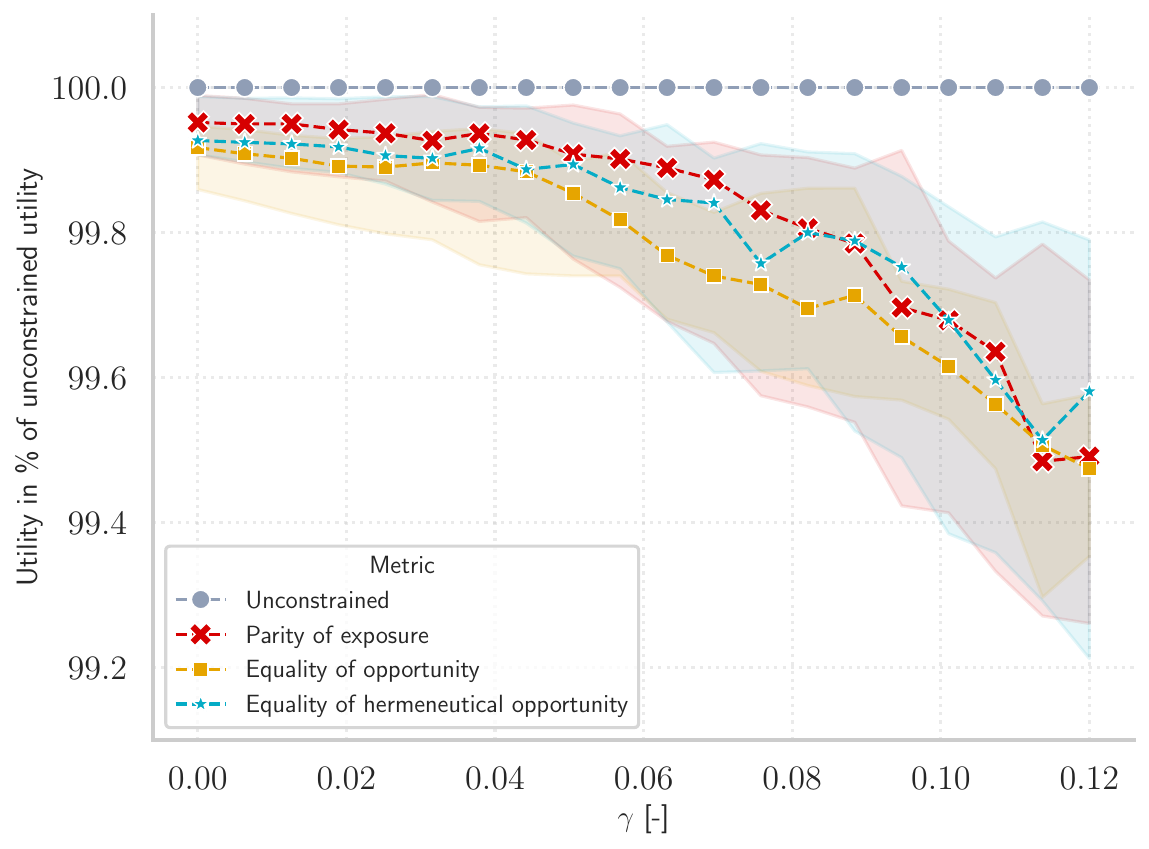}
  \caption{Utility (\% of unconstrained).}
  \label{fig:sim_gamma_utility_app}
\end{subfigure}\hspace*{.75cm}
\begin{subfigure}[t]{0.4\textwidth}
  \centering
  \includegraphics[width=\linewidth]{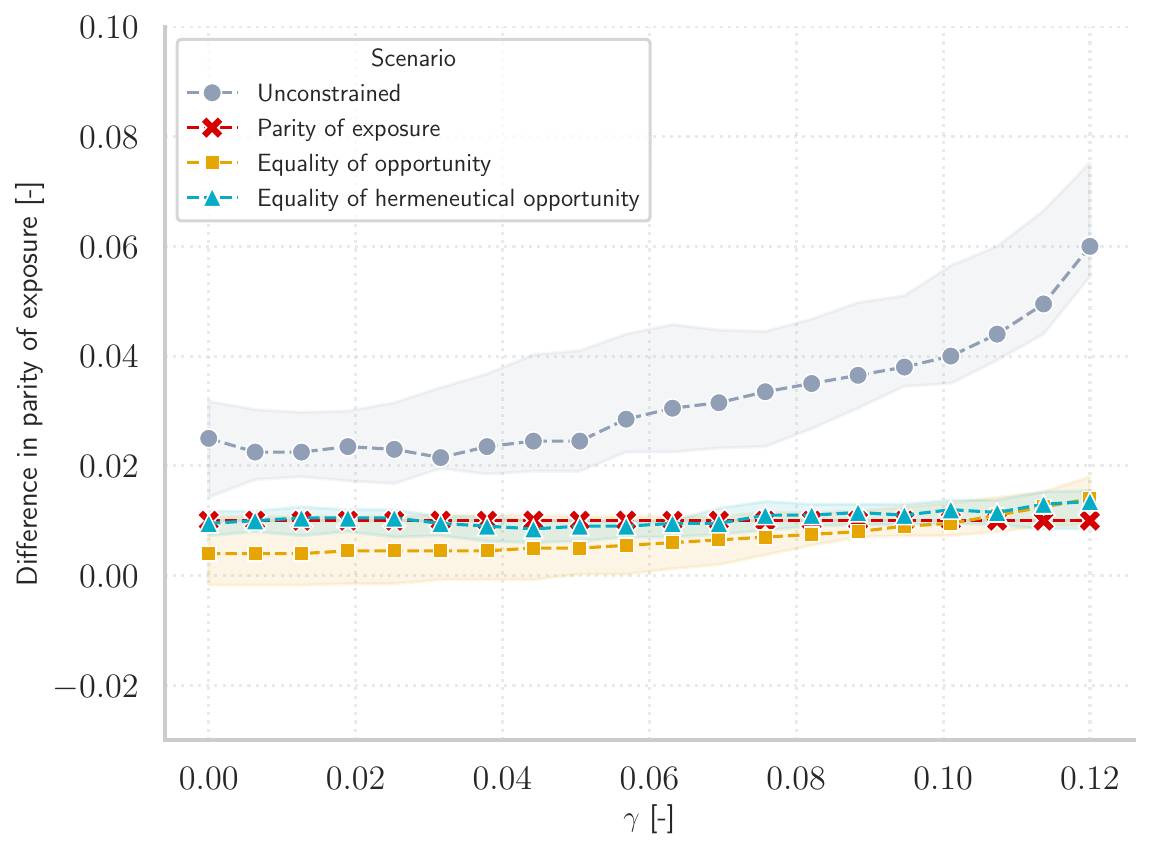}
  \caption{Parity of exposure difference.}
  \label{fig:sim_gamma_parity_app}
\end{subfigure}
\caption{\textbf{Additional simulation results with varying hermeneutical-cost weight $\gamma$.} Achieved utility (as \% of the unconstrained solution) and parity of exposure difference.}
\label{fig:sim_additional_diagnostics}
\end{figure*}

We here report a set of results that complement the main analysis. First, we fix $\gamma=0$ and we vary $\beta_b$, thus considering the same setting as in \cite{baumann2024fairness}, to show the consistency of our results with what we regard as the baseline for our framework. Second, we vary $\gamma$ (i.e., the hermeneutical-cost weight) while keeping all other parameters fixed. This second scenario allows us to examine how increasing the weight on the hermeneutical-cost term affects the achieved utility and exposure disparities.

When looking at \figurename{~\ref{fig:baseline}}, we achieve results that are comparable with the one shown in \cite{baumann2024fairness}. In particular, the constrained policies exhibit the largest utility loss (relative to the unconstrained benchmark) for intermediate values of $\beta_b$ (around $\beta_b \approx 0.2$-$0.3$), with smaller losses toward both ends of the explored range. Moreover, \figurename{~\ref{fig:sim_beta_parity_app}} shows that, while the unconstrained policy yields a large positive parity of exposure difference quickly rising, the three fairness constraints variants keep the parity of exposure difference small and nearly flat across $\beta_b$.

Meanwhile, \figurename{~\ref{fig:sim_additional_diagnostics}} shows the results achieved when varying $\gamma$. As expected, \figurename{~\ref{fig:sim_gamma_utility_app}} shows that achieved utility under all three fairness constraint variants declines overall as $\gamma$ increases, with the gap between constrained and unconstrained setting widening at higher $\gamma$. In terms of exposure disparities, \figurename{~\ref{fig:sim_gamma_parity_app}} indicates that the unconstrained policy produces a growing parity of exposure difference as $\gamma$ increases, whereas the constrained policies keep the parity gap low with only minor variation across the full range of $\gamma$. Taken together, these patterns indicate that placing greater weight on the hermeneutical-cost term is associated with a larger utility penalty for satisfying distributive and hermeneutical constraints, and increased exposure inequality when no distributive constraint is imposed. This effect is attenuated once distributive and hermeneutical constraints are imposed.

\subsection{Additional Uptake Configurations}\label{app:b_1}
\begin{figure}[t]
    \centering
    \includegraphics[scale=.30]{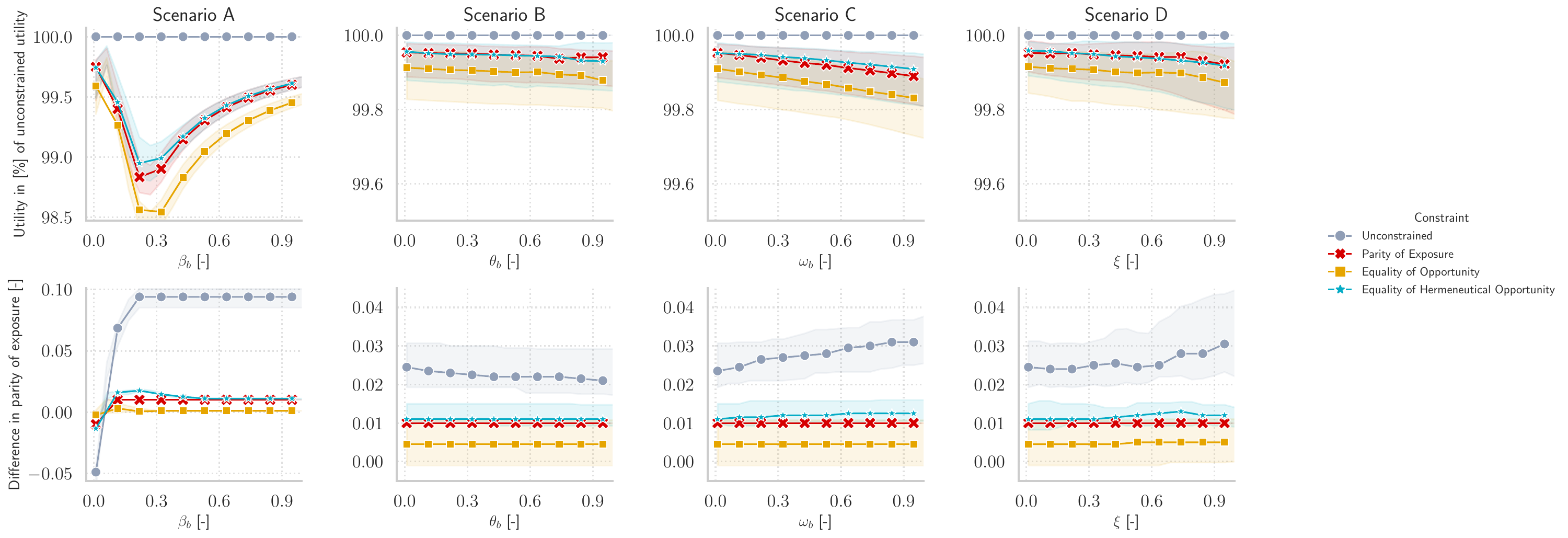}
    \caption{\textbf{Simulation results for Scenarios A--D (Table~\ref{tab:parameters}) under an $a$-advantaged uptake configuration.} Uptake probabilities are sampled as $\rho_a \sim \mathrm{Beta}(8,2)$ and $\rho_b \sim \mathrm{Beta}(3,7)$. Top row: achieved utility (in \% of the unconstrained solution). Bottom row: parity of exposure difference. Solid lines show the median outcomes across the 100 replications, while the shaded bands show their range according to the 0.25-0.75 quartiles.}
    \label{fig:scenario_1_new}
\end{figure}

\begin{figure}[t]
    \centering
    \includegraphics[scale=.30]{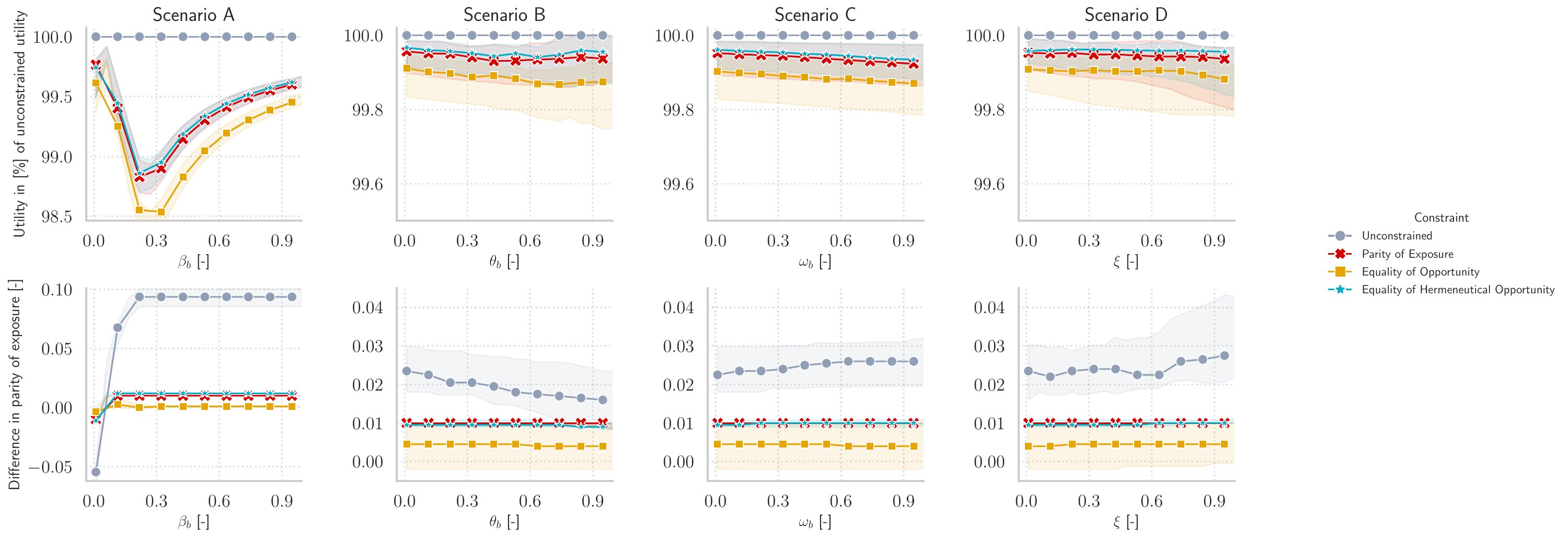}
    \caption{\textbf{Simulation results for Scenarios A--D (Table~\ref{tab:parameters}) under a neutral high-uptake configuration.} Uptake probabilities are sampled identically across groups, with $\rho_a=\rho_b \sim \mathrm{Beta}(7,3)$. Top row: achieved utility (in \% of the unconstrained solution). Bottom row: parity of exposure difference. Solid lines show the median outcomes across the 100 replications, while the shaded bands show their range according to the 0.25-0.75 quartiles.}
    \label{fig:scenario_2_new}
\end{figure}

\begin{figure}[t]
    \centering
    \includegraphics[scale=.30]{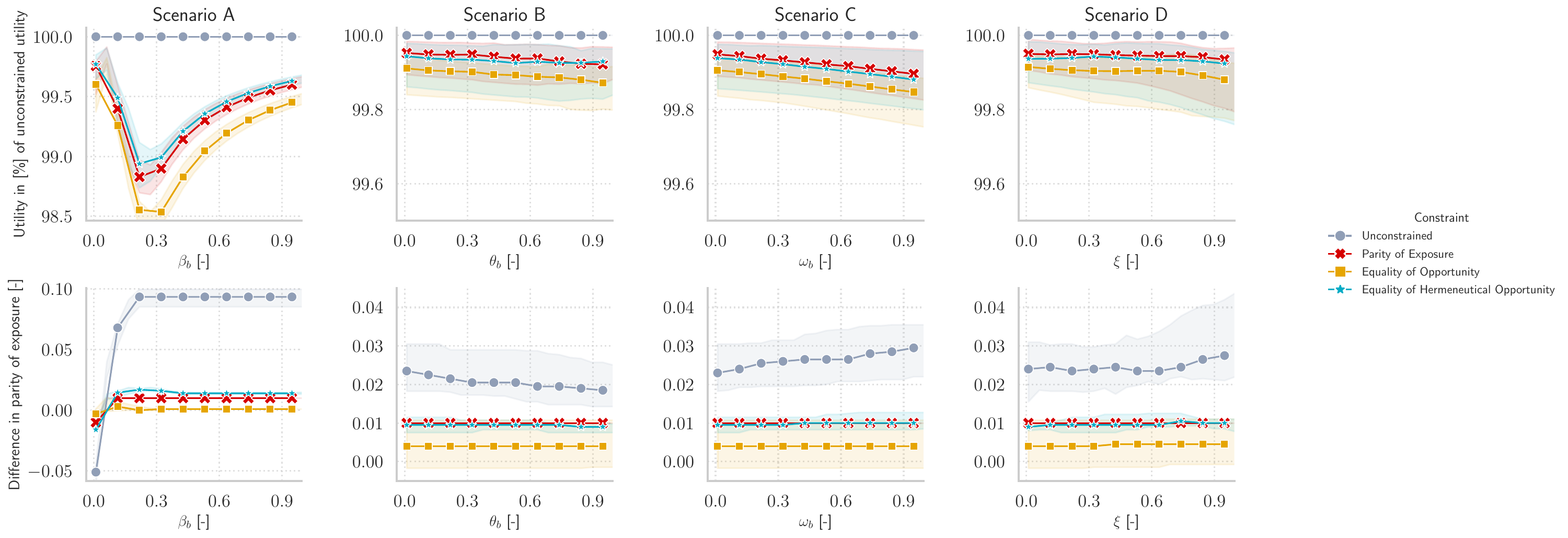}
    \caption{\textbf{Simulation results for Scenarios A--D (Table~\ref{tab:parameters}) under a neutral lower-uptake configuration.} Uptake probabilities are sampled identically across groups, with $\rho_a=\rho_b \sim \mathrm{Beta}(4,6)$. Top row: achieved utility (in \% of the unconstrained solution). Bottom row: parity of exposure difference. Solid lines show the median outcomes across the 100 replications, while the shaded bands show their range according to the 0.25-0.75 quartiles.}
    \label{fig:scenario_3_new}
\end{figure}

To assess how sensitive the qualitative patterns are to the distribution of uptake probabilities across groups, we additionally report results for the three alternative uptake configurations introduced in Section \ref{sec:6}. In the main text, results are shown for the $b$-advantaged case (with $\rho_a \sim \mathrm{Beta}(4,6)$ and $\rho_b \sim \mathrm{Beta}(7,3)$). Here, we complement that analysis with three additional configurations: (i) an $a$-advantaged case with $\rho_a \sim \mathrm{Beta}(8,2)$ and $\rho_b \sim \mathrm{Beta}(3,7)$ (Figure \ref{fig:scenario_1_new}); (ii) a neutral high-uptake configuration with $\rho_a=\rho_b \sim \mathrm{Beta}(7,3)$ (Figure \ref{fig:scenario_2_new}); (iii) a neutral lower-uptake configuration with $\rho_a=\rho_b \sim \mathrm{Beta}(4,6)$ (Figure \ref{fig:scenario_3_new}). All other simulation settings are unchanged.

Across these configurations, the main qualitative patterns remain stable. In particular, constrained solutions continue to achieve utility levels close to the unconstrained benchmark, while keeping exposure disparities substantially smaller than in the unconstrained case. Changing the uptake distributions does affect the magnitude of the unconstrained exposure gap, but does not alter the broader comparative pattern across constraints. Overall, these additional scenarios suggest that the main simulation conclusions are not driven by a single parameterization of the $\mathrm{Beta}$ distributions, although the size of the resulting disparities remains sensitive to the modeled uptake configuration.

\section{Protocol sketch for estimating uptake probability}\label{app:c}

\begin{figure}[t]
\centering

\definecolor{startend}{RGB}{255,255,209}
\definecolor{decisiongreen}{RGB}{195,252,181}
\definecolor{expone}{RGB}{247,206,205}
\definecolor{exptwo}{RGB}{242,167,162}
\definecolor{expthree}{RGB}{239,137,133}
\definecolor{processbeige}{RGB}{249,217,181}
\definecolor{outcomepurple}{RGB}{176,179,249}

\resizebox{\linewidth}{!}{
\begin{tikzpicture}[
    font=\scriptsize,
    >={Latex[length=1.8mm]},
    line width=0.7pt,
    every node/.style={align=center},
    redbox/.style={
        draw,
        rounded corners=3pt,
        minimum height=0.78cm,
        text width=2.55cm,
        inner sep=2pt
    },
    purplebox/.style={
        draw,
        rounded corners=3pt,
        minimum height=0.78cm,
        text width=2.25cm,
        inner sep=2pt
    }
]

\node[
    draw,
    ellipse,
    fill=startend,
    minimum width=1.6cm,
    minimum height=0.72cm
] (start) at (0,0) {start};

\node[
    draw,
    diamond,
    aspect=2.1,
    fill=decisiongreen,
    minimum width=2.9cm,
    minimum height=1.05cm,
    inner sep=1pt
] (assign) at (3.0,0) {attribute-myopic\\assignment};

\node[redbox, fill=expone]   (none) at (6.9, 1.65) {no exposure};
\node[redbox, fill=exptwo]   (low)  at (6.9, 0.00) {low exposure\\neutral/skewed};
\node[redbox, fill=expthree] (rep)  at (6.9,-1.65) {repeated exposure\\neutral/skewed};

\node[
    draw,
    fill=processbeige,
    minimum width=2.35cm,
    minimum height=0.78cm
] (eval) at (10.9,0) {uptake\\evaluation};

\node[purplebox, fill=outcomepurple] (understanding) at (14.9, 1.65) {understanding};
\node[purplebox, fill=outcomepurple] (application)   at (14.9, 0.00) {scenario\\application};
\node[purplebox, fill=outcomepurple] (teachback)     at (14.9,-1.65) {teach-back};

\node[
    draw,
    fill=processbeige,
    minimum width=2.3cm,
    minimum height=0.78cm
] (aggregate) at (18.9,0) {aggregate \&\\normalize};

\node[
    draw,
    ellipse,
    fill=startend,
    minimum width=1.95cm,
    minimum height=0.72cm
] (rho) at (22.2,0) {estimate\\$\hat{\rho}_x$};

\coordinate (Lsplit) at (4.9,0);
\coordinate (Ltop)   at (4.9,1.65);
\coordinate (Lbot)   at (4.9,-1.65);

\coordinate (Rmerge) at (8.9,0);
\coordinate (Rtop)   at (8.9,1.65);
\coordinate (Rbot)   at (8.9,-1.65);

\coordinate (Esplit) at (13.0,0);
\coordinate (Etop)   at (13.0,1.65);
\coordinate (Ebot)   at (13.0,-1.65);

\coordinate (Gmerge) at (16.8,0);
\coordinate (Gtop)   at (16.8,1.65);
\coordinate (Gbot)   at (16.8,-1.65);

\draw[->] (start.east) -- (assign.west);

\draw (assign.east) -- (Lsplit);

\draw (Ltop) -- (Lbot);
\draw (Lsplit) -- (Ltop);
\draw (Lsplit) -- (Lbot);

\draw[->] (Ltop) -- (none.west);
\draw[->] (Lsplit) -- (low.west);
\draw[->] (Lbot) -- (rep.west);

\draw (none.east) -- (Rtop);
\draw (low.east)  -- (Rmerge);
\draw (rep.east)  -- (Rbot);

\draw (Rtop) -- (Rbot);
\draw[->] (Rmerge) -- (eval.west);

\draw (eval.east) -- (Esplit);

\draw (Etop) -- (Ebot);
\draw (Esplit) -- (Etop);
\draw (Esplit) -- (Ebot);

\draw[->] (Etop) -- (understanding.west);
\draw[->] (Esplit) -- (application.west);
\draw[->] (Ebot) -- (teachback.west);

\draw (understanding.east) -- (Gtop);
\draw (application.east)   -- (Gmerge);
\draw (teachback.east)     -- (Gbot);

\draw (Gtop) -- (Gbot);
\draw[->] (Gmerge) -- (aggregate.west);

\draw[->] (aggregate.east) -- (rho.west);

\end{tikzpicture}
}

\caption{Individual-level pipeline for estimating uptake probability $\rho_x$.}
\label{fig:rho_protocol}
\end{figure}
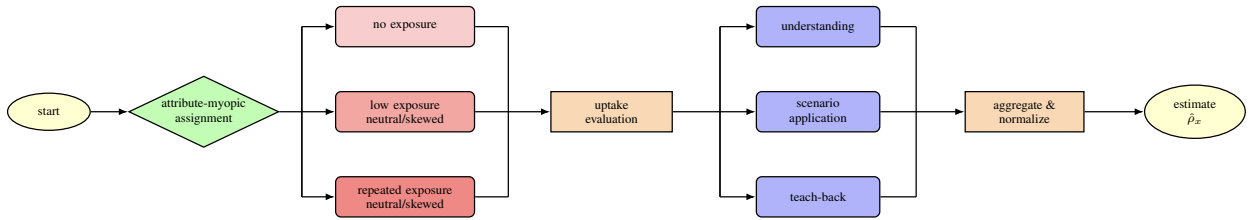

A concrete path toward empirical validation of our framework is a controlled ad-delivery study designed to estimate individual-level uptake probability $\rho_x$, a pipeline of which is sketched in  Figure \ref{fig:rho_protocol}.

At a high level, each participant first enters the experiment and is assigned to an attribute-myopic delivery condition. This places the participant into one of three exposure conditions: no exposure, low exposure with neutral or skewed framing, or repeated exposure with neutral or skewed framing. After exposure, the participant completes a post-exposure uptake evaluation. This evaluation is intended to measure understanding and combines three components: content understanding, scenario-based content application, and a teach-back or paraphrase task \cite{baseman2013public,talevski2020teach}. For instance, the instrument may ask participants to explain the main message of the ad, apply it to a hypothetical case, and restate it in their own words. The resulting responses are then aggregated and normalized into an individual uptake score, which yields an estimate $\hat{\rho}_x$ for that participant.

At minimum, estimating uptake $\rho_x$ requires linked data on participant features, assigned ad condition, ad variant or framing, exposure count, and post-exposure uptake responses. Estimating platform-side utility parameters further requires standard internal delivery variables such as bid value and predicted click probability. Enforcing or auditing the group-level fairness constraints additionally requires access to protected-group labels or fairness-relevant proxies. This highlights an implementation boundary. The individual-level hermeneutical cost can in principle be modeled without direct access to protected attributes, whereas the group-level fairness constraints cannot be imposed or evaluated without some group information. In practice, platforms may infer uptake-related signals from behavioral traces or learned user representations, but such proxies may encode behavioral bias, unequal access, or platform-specific engagement artifacts. As a result, optimizing on poorly validated proxies could amplify interpretative distortions while appearing well calibrated under the model. When protected attributes are unavailable or highly regulated, the framework may still support individual-level robustness checks and sensitivity analysis, but not the full enforcement of the fairness constraints introduced in Section~\ref{sec:4}.

\paragraph{Remark on parameter calibration.} The model parameters play different operational roles. The economic parameters $\alpha$ and $\beta$ are platform-side utility quantities and can in principle be obtained from delivery logs or advertiser bids. By contrast, the epistemic parameters $\theta$, $\omega$, and $\xi$ should be interpreted as decision weights encoding the costs of successful uptake, failed uptake, and exclusion. For this reason, the empirical goal is not necessarily to identify them pointwise, but to calibrate or bound them. Once $\rho_x$ has been estimated, variation across the three exposure conditions can be used to inform these remaining parameters. Comparisons between no exposure and exposure conditions can help bound exclusion-related costs, while comparisons between lower-uptake and repeated-exposure conditions can inform the calibration of penalties associated with failed or distorted uptake. These comparisons are therefore best understood as inputs to calibration and sensitivity analysis, rather than as a strategy for point identification.

\end{document}